\documentclass[manuscript,screen]{acmart}

\usepackage{pstricks}
\usepackage{verbatim,framed}
\usepackage{amsmath,amsthm}
\usepackage{xcolor}
\theoremstyle{remark}
\newtheorem{rem}{Remark}
\AtBeginDocument{%
  }

\setcopyright{acmlicensed}
\copyrightyear{2018}
\acmYear{2018}
\acmDOI{XXXXXXX.XXXXXXX}

\acmISBN{978-1-4503-XXXX-X/18/06}

\usepackage[american,cuteinductors,siunitx]{circuitikz}
\usetikzlibrary{calc}
\ctikzset{bipoles/thickness=1}
\ctikzset{bipoles/length=0.8cm}

\definecolor{light-gray}{gray}{0.95}
\newenvironment{colorverbatim}[1][light-gray]%
{%
	\definecolor{shadecolor}{named}{#1}%
	\topsep=0ex\relax
	\shaded
	\verbatim
}%
{%
	\endverbatim
	\endshaded
}%

\usepackage{bm}

\begin{document}

\title{Symbolic and User-friendly Geometric Algebra Routines ($\mathtt{SUGAR}$) for Computations in Matlab}

\author{Manel Velasco}
\email{manel.velasco@upc.edu}
\orcid{0000-0002-0764-3063}
\affiliation{%
	\institution{Automatic Control Department, Universitat Politècnica de Catalunya (UPC)}
	\city{Barcelona}
	\country{Spain}
	\postcode{08028}
}
\author{Isiah Zaplana}
\email{isiah.zaplana@upc.edu}
\orcid{0000-0002-0862-3240}
\author{Arnau Dòria-Cerezo}
\email{arnau.doria@upc.edu}
\orcid{0000-0001-9352-066X}
\affiliation{%
	\institution{Institute of Industrial and Control Engineering, Universitat Politècnica de Catalunya (UPC)}
	\city{Barcelona}
	\country{Spain}
	\postcode{08028}
}
\author{Pau Martí}
\email{pau.marti@upc.edu}
\orcid{0000-0002-5189-0782}
\affiliation{%
	\institution{Automatic Control Department, Universitat Politècnica de Catalunya (UPC)}
	\city{Barcelona}
	\country{Spain}
	\postcode{08028}
}

\renewcommand{\shortauthors}{Velasco et al.}

\begin{abstract}
  Geometric Algebra (GA) provides a unified, compact mathematical framework for geometric computing, simplifying relations typically handled with more complex tools like matrix multiplication. In fields like robotics, GA replaces conventional coordinate-based approaches with the multiplication of special elements called rotors, offering greater efficiency. Despite its advantages, GA’s complexity and the lack of symbolic tools hinder its broader adoption among applied mathematicians and engineers. To address this, this paper introduces SUGAR (Symbolic and User-friendly Geometric Algebra Routines), an open-source Matlab toolbox. SUGAR streamlines GA usage in Matlab through a collection of user-friendly functions that support both numeric and symbolic computations, even in high-dimensional algebras. Designed for applied mathematics and engineering, it enables intuitive manipulation of geometric elements and transformations in two- and three-dimensional projective and conformal GAs, consistent with established computational methods. Moreover, SUGAR manages multivector functions such as exponential, logarithmic, sinusoidal, and cosine operations, enhancing its applicability in domains like robotics, control systems, and power electronics. Finally, this paper also presents three validation examples across these fields, showcasing SUGAR's practical utility in solving real-world engineering and applied mathematics problems.
\end{abstract}

\keywords{Geometric Algebra, Symbolic Computations, High-Dimensional Systems, Matlab, Engineering}

\received{24 March 2024}
\received[revised]{XXX}
\received[accepted]{XXX}

\maketitle

\section{Introduction}\label{s:int}

From an applied mathematics and engineering perspective, geometric algebra (GA) \cite{Hes84} can be described as a mathematical tool for geometric computing, as it provides a framework that allows a unified and compact approach to geometric relations which in other mathematical systems are typically described using different elements. An illustration of this occurs in the field of robotics \cite{Bay20}, where the end-effector's position is a Euclidean point, while its orientation is represented by a quaternion. In GA, however, both can be represented as a single element, referred to as a multivector. Projective or Plane-based Geometric Algebra (PGA) and Conformal Geometric Algebra (CGA) \cite{Hes01} introduce an even more elegant and intuitive approach to geometry by encoding both the geometric entities (such as points, lines, planes and spheres) and geometric transformations (such as rotations and translations) as elements of the algebra, which in turn allows one to operate with them as one does with real numbers. Beyond applied mathematics and engineering, GA also leads to the simplification of many otherwise complex equations, making them more intuitive and easy to handle. An example of this is seen in the well-known Maxwell equations, which can be reduced to a single equation using GA \cite{Cha14}.

The inherent geometric intuition and the potential for simplifying complex equations make GA relevant in various engineering fields. Disciplines with a predominant geometric interpretation like robotics already apply GA, PGA and CGA either in kinematics, dynamics, or tracking and control of robotic systems \cite{Bay20, Lav18, Bay22, Tob23}. In addition, since applied mathematicians and engineers are always interested in keeping equations as simple and compact as possible, the above-mentioned inherent simplifying feature of GA has attracted attention from other disciplines as well. This attraction has been discussed in several surveys and tutorials. In particular, various studies have explored GA applications in fields such as signal and image processing, computer vision, and artificial intelligence \cite{Hit13,Bay21,Hit24}, hile others have focused on its use in electrical and electronic engineering \cite{Cha14}.

This interest on GA, PGA and CGA triggers the need for software implementations where mathematicians and engineers may feel more comfortable. Motivated by this need, this paper presents an implementation of GA, PGA and CGA for~Matlab\textsuperscript\textregistered\footnote{\url{https://www.mathworks.com}}.  Matlab is a proprietary software system developed and sold by The MathWorks. It has been commercially available since 1984 and it is now considered as a standard tool at most universities and industries worldwide. Matlab is an interactive system whose basic data element is an {\it array} that does not require dimensioning, easing the calculus  with matrices of real and complex numbers. Specific application domains are collected in packages referred to as {\it toolboxes}, covering symbolic computation, control theory, simulation, optimization, and several other fields of applied science and engineering. 

The GA, PGA, and CGA implementation presented in this paper is given under the $\mathtt{SUGAR}$ toolbox \cite{SUGAR}, that stands for Symbolic and User-friendly Geometric Algebra Routines. The name aims to stress two of SUGAR's key features. The first one is its capability to allow symbolic computations, scaling to any (high) dimension (where the limitation is the CPU power, but not the underlying algorithmics). The second one refers to the fact that SUGAR has been developed to offer a more natural language that can be directly applied to complex problems in both applied mathematics and engineering, thus easing its usage and speeding-up its learning curve, that is, being user-friendly.

It is interesting to note that $\mathtt{SUGAR}$  was born  for covering the specific research needs faced in the modeling, analysis and control of (unbalanced) three-phase electrical systems \cite{Vel23}. However, the rapid interest on the few original routines that was shown by GA-research colleagues triggered a new coding effort that lead to the current version of $\mathtt{SUGAR}$, serving different fields, as illustrated in the application examples presented at the end of the present article. 

The rest of this manuscript is organized as follows. First, Section~\ref{s:ga} provides a brief introduction to geometric algebra, as well as projective and conformal geometric algebras, to establish the foundation for understanding the rest of the manuscript. Section~\ref{s:soa} presents a non-exhaustive state-of-the-art analysis of available geometric algebra-related software implementations. Then, Section~\ref{s:des} describes the main features of $\mathtt{SUGAR}$, as well as the initial steps to utilize these features. Finally, Section~\ref{s:app} presents application examples, and Section~\ref{s:con} concludes the paper.

\section{A primer on GA}\label{s:ga}
In this section, a brief introduction to GA, PGA and CGA is provided. Readers interested in a more detailed treatment of the subject are referred to the classical texts \cite{Hes84,Hes01,Doran03,Dor22}.

\subsection{Geometric algebra}
Let $\mathbb{R}^{p,q,r}$ be a the pseudo-Euclidean space with orthonormal basis $\{e_{1},\dots,e_{p},e_{p+1},\dots,e_{p+q},e_{p+q+1},\dots,e_{p+q+r}\}$, where the basis elements $\{e_1,\dots,e_p\}$ square to $1$, the basis elements $\{e_{p+1},\dots,e_{p+q}\}$ square to $-1$, and the basis elements $\{e_{p+q+1},\dots,e_{p+q+r}\}$ square to $0$. The geometric algebra of $\mathbb{R}^{p,q,r}$, denoted by $\mathcal{G}_{p,q,r}$, is a vector space where the operations defined in $\mathbb{R}^{p,q,r}$, i.e., the addition and multiplication by scalars, are extended naturally. An additional operation, the geometric product, is defined, and acts on vectors of the algebra as follows: 
\begin{equation}\label{geometricproduct}
	v_{1}v_{2} = v_{1}\cdot v_{2} + v_{1}\wedge v_{2},\,\text{ for } v_{1},v_{2}\in\mathbb{R}^{p,q,r},
\end{equation}
where $\cdot$ denotes the inner or dot product and $\wedge$ denotes the outer product. 

The outer product is anticommutative, which implies that for any vector $v$, $v\wedge v = 0$. In this case, the geometric product simplifies to $vv = v\cdot v$ which is also denoted as $v^2$. The outer product of two distinct vectors $v_{1},v_{2}$ is a new element of $\mathcal{G}_{p,q,r}$, which is termed a bivector, is said to have grade two and is denoted by $v_{1}\wedge v_{2}$. By extension, the outer product of a bivector with a vector is known as a trivector and is denoted by $(v_{1}\wedge v_{2})\wedge v_{3}$. Clearly, trivectors have grade three. This can be generalized to an arbitrary dimension. Thus,
\begin{equation}\label{rblade}
	(v_{1}\wedge v_{2} \wedge \dots \wedge v_{k-1})\wedge v_{k}
\end{equation}
denotes an \textit{$k$-blade}, an element of $\mathcal{G}_{p,q,r}$ with grade $k$. 

A bivector $v_{1}\wedge v_{2}$ can be interpreted as the oriented area defined by the vectors $v_{1}$ and $v_{2}$. Thus, $v_{2}\wedge v_{1}$ has opposite orientation, which illustrates the anticommutativity of the outer product. Analogously, a trivector is interpreted as the oriented volume defined by its three composing vectors. Since the volume generated by $(v_{1}\wedge v_{2})\wedge v_{3}$ is the same as the volume generated by $v_{1}\wedge (v_{2}\wedge v_{3})$, it is also deduced that the outer product is associative. Therefore, $k$-blades can be denoted simply as:
\begin{equation}\label{ibladebis}
	v_{1}\wedge v_{2} \wedge \dots \wedge v_{k}.
\end{equation}
Linear combinations of $k$-blades are known as $k$-vectors, while linear combinations of $k$-vectors, for $0\leq k\leq p+q+r$, are called multivectors. Multivectors are the most important elements of geometric algebra.

Applied to the basis elements $\{e_i\}$, the geometric product acts as follows:
\begin{equation}\label{elementsofbasis}
	e_{i}e_{j}  = \left\{\begin{array}{lcl}
		1 & \text{for} & i=j\text{ and }i\leq p\\
		-1 & \text{for} & i=j\text{ and } p< i\leq p+q\\
		0 & \text{for} & i=j\text{ and } p+q<i\leq p+q+r\\
		e_{i}\wedge e_{j} & \text{for} & i\neq j
	\end{array}\right.
\end{equation} 
Then, $\{e_{1},\dots,e_{p},e_{p+1},\dots,e_{p+q},e_{p+q+1},\dots,e_{p+q+r}\}$ can be expanded to a basis of $\mathcal{G}_{p,q,r}$ that contains, for each $0\leq k\leq p+q+r$, $C(p+q+r,k)$ grade $k$ elements, where $C(p+q+r,k)$ represents the combinatorial number of $p+q+r$ over $k$:
\begin{equation}\label{basisGn}
	\begin{split}
		&\text{Scalar: } 1\\
		&\text{Vectors: } e_{1},\dots,e_{p+q+r}\\
		&\text{Bivectors: } \{e_{i}\wedge e_{j}\}_{1\leq i<j\leq p+q+r}\\
		&\text{Trivectors: } \{e_{i}\wedge e_{j}\wedge e_{k}\}_{1\leq i<j<k\leq p+q+r}\\
		&\vdots\\
		&k-\text{vectors: } \{e_{i_{1}}\wedge\dots\wedge e_{i_{k}}\}_{1\leq i_{1}<\dots<i_{k}\leq n}\\
		&\vdots\\
		&\text{$p+q+r$-blade: } e_{1}\wedge\dots\wedge e_{p+q+r}
	\end{split}
\end{equation}
which sums up to total of $2^{p+q+r}$ elements. Understanding how the geometric product acts on the basis elements of $\mathcal{G}_{p,q,r}$ allows for its extension to arbitrary multivectors. The grade $p+q+r$ element $e_{1}\wedge\dots\wedge e_{p+q+r}$ is known as the pseudoscalar and is usually denoted by $I$. If $r=0$, pseudoscalars allow for the definition of one of the main operators of geometric algebra, the dual operator. Its action over an $k$-vector $A_{k}$ is:
\begin{equation}\label{dual}
	A_{k}^{\ast} = A_kI,
\end{equation}
where $A_{k}^{\ast}$ is an $(p+q-k)$-vector. In particular, for two multivectors $A$ and $B$, the following identity holds:
\begin{equation}\label{deMorgan}
	(A\wedge B)^\ast = A\cdot B^\ast.
\end{equation}
The geometric algebras with $p=n$ and $q=r=0$ are the algebras over an $n$-dimensional Euclidean space $\mathbb{R}^n$. These are denoted $\mathcal{G}_n$ for $n\in\mathbb{N}$. The bivectors of $\mathcal{G}_n$ play an important role since they can be used to describe $n$-dimensional rotations. Indeed, in geometric algebra, rotations are described using rotors. If a point $x\in\mathbb{R}^{n}$ is rotated by an angle $\theta$ around an axis $\ell$, the rotor $R$ defining such a rotation in $\mathcal{G}_n$ is:
\begin{equation}\label{RotorDescrip}
	R = e^{-\frac{\theta}{2}B} = \cos\biggl(\dfrac{\theta}{2}\biggr)-\sin\biggl(\dfrac{\theta}{2}\biggr)B,
\end{equation}
where $B$ is the unit bivector, i.e., $B^2 = -1$, representing the hyperplane normal to $\ell$. The second identity is obtained by expanding the Taylor series of the exponential $e^{-\frac{\theta}{2}B}$ and regrouping terms \cite{Doran03}. The rotated point $x'$ is calculated by sandwiching $x$ between $R$ and its reverse $\widetilde{R}$:
\begin{equation}\label{sandwiching}
	x' = Rx\widetilde{R}
\end{equation}
where
\begin{equation}
	\widetilde{R} = \cos\biggl(\dfrac{\theta}{2}\biggr)+\sin\biggl(\dfrac{\theta}{2}\biggr)B
\end{equation}
and $R\widetilde{R}=1$. Furthermore, equation (\ref{sandwiching}) can be extended to arbitrary multivectors. In addition, to specify what unit multivectors are, a norm is defined in $\mathcal{G}_n$. For an arbitrary multivector $A$, the norm is:
\begin{equation}
	\|A\| = \sqrt{\left<A\widetilde{A}\right>_0},
\end{equation}
where $\left<\cdot\right>_0$ is the grade-0 projection operator, i.e., $\left<\cdot\right>_0$ extracts the scalar elements of the argument multivector. In fact, grade-$k$ projection operators constitute an important family of linear operators in $\mathcal{G}_{n}$. They are denoted by $\left<\cdot\right>_{k}$ for $0\leq k\leq n$. When applied to an arbitrary multivector $A$, $\left<A\right>_{k}$ projects onto the grade-$k$ components in $A$, i.e., $\left<A\right>_{k}$ returns the components of $A$ that can be expressed as a linear combination of $\{e_{i_{1}}\wedge\dots\wedge e_{i_{k}}\}_{1\leq i_{1}<\dots<i_{k}\leq n}$. Obviously, if $A_{k}$ denotes a $k$-vector, then $\left<A_{k}\right>_{k} = A_{k}$.

Using these operators, general multivectors $A\in\mathcal{G}_{n}$ can be expressed as:
\begin{equation}\label{multivectors}
	A = \left<A\right>_{0} + \left<A\right>_{1} + \dots + \left<A\right>_{n}.
\end{equation}
Hence, the set of all $k$-vectors for a given $1\leq k\leq n$ is a vector subspace of $\mathcal{G}_{n}$ denoted by $\left<\mathcal{G}_{n}\right>_{k}$ and spanned by $B_{k} = \{e_{i_{1}}\wedge\dots\wedge e_{i_{k}}\}_{1\leq i_{1}<\dots<i_{k}\leq n}$.

The multivector representation (\ref{multivectors}) is useful in defining another important operator in $\mathcal{G}_{n}$. This linear operator is known as the \textit{reversion operator} and is denoted by the superscript $\sim$. The reversion is defined over the geometric product of $m$ vectors as:
\begin{equation}\label{reverse}
	(a_{1}\cdots a_{m})^{\sim} = a_{m}\cdots a_{1}.
\end{equation}
Applied to $k$-vectors:
\begin{equation}\label{reverseblades}
	\widetilde{A}_k = (-1)^{\frac{k(k-1)}{2}}A_{k}
\end{equation}
due to the anticommutativity of the outer product. Finally, since reversion is a linear operator, the reverse of an arbitrary multivector is:
\begin{equation}\label{reversemultivector}
	\widetilde{A} = \left<\widetilde{A}\right>_{0} + \dots + \left<\widetilde{A}\right>_{n} = \left<A\right>_{0} + \left<A\right>_{1} - \left<A\right>_{2} + \dots + (-1)^{\frac{n(n-1)}{2}}\left<A\right>_{n}.
\end{equation}

\subsection{Projective geometric algebra}\label{PGA}

The projective or plane-based model extends the $n$-dimensional Euclidean space $\mathbb{R}^n$ by adding an extra basis vector $e$ with the property $e^2 = 0$, i.e., vector $e$ is a null vector. This null vector is associated with the point at infinity (and this is why this geometric algebra is known as the projective geometric algebra).

The projective or plane-based geometric algebra (PGA) of $\mathbb{R}^{n}$, denoted $\mathcal{G}_{n,0,1}$, can be seen as the geometric algebra of $\mathbb{R}^{n,0,1}$. One of the key features of PGA is that it allows the encoding of translations as rotors. Thus, all proper rigid body transformations are represented by the same structure within the algebra. This property enables the use of PGA in kinematic-related problems, such as in robotics. In particular, the addition of the null basis vector $e$ allows for the definition of null bivectors of the form $B = b \wedge e$, where $b$ is a non-null vector of $\mathbb{R}^n$. These bivectors simplify the Taylor expansion of $e^{-\frac{\theta}{2}B}$ to a single linear term of the form:
\begin{equation}
	e^{-\frac{\theta}{2}B} = 1 - \dfrac{b\wedge e}{2}
\end{equation}
which encodes a rotation around a line with one point at infinity, i.e., a translation along the direction vector $b\in\mathbb{R}^{n}$. These rotors are denoted as $T_b$, and they are applied to other vectors using the sandwich product, similar to any other rotor. In other words, to translate vector $x$ along the direction $b$:
\begin{equation}\label{translator_PGA}
	x' = T_b x\widetilde{T}_b,
\end{equation}
where $x'$ is the translated vector and $\widetilde{T} = 1 + \frac{b\wedge e}{2}$.

In addition, geometric entities such as planes, lines, and points can be represented as elements of the algebra. Since the key elements in this algebra are planes (hence its name), these geometric entities can be expressed in terms of planes or their constituting elements. In particular:
\begin{itemize} 
	\item $\bm{\pi} = n - \delta e$ is a vector representing a plane $\pi$, where $n$ denotes the unit vector normal to the plane and $\delta$ its orthogonal distance to the origin. \item $\bm{\ell} = \bm{\pi}_1\wedge \bm{\pi}_2$ is a bivector representing the line resulting from the intersection of the planes represented by $\bm{\pi}_1$ and $\bm{\pi}_2$, if such a line exists. This line can also be represented using its constituting elements, i.e., its direction vector $v$ and a point $p$ lying on it: $\bm{\ell}= ve_{123} - e(p\wedge v)e_{123}$. 
	\item $\bm{x} = \bm{\pi}_1\wedge\bm{\pi}_2\wedge\bm{\pi}_3$ is a trivector representing the point resulting from the intersection of the planes represented by $\bm{\pi}_1$, $\bm{\pi}_2$, and $\bm{\pi}_3$, if such a point exists. This point can also be represented using its Euclidean coordinates $x$: $\bm{x}= (1-ex)e_{123}$. 
\end{itemize} 
Given a rotor $T_b$ as in equation \eqref{translator_PGA}, one can translate any of these geometric representations along the direction $b$ in the same way as with vectors.

Finally, the pseudoscalar for this algebra is given by $I = e e_1e_2\dots e_n$. However, due to the presence of the null vector $e$, it satisfies $I^2 = 0$. As a result, this pseudoscalar is not suitable for computing the dual in the usual manner, i.e., as introduced in Section 2.1. Instead, duality in PGA is performed using the Hodge star operator $\star$. When applied to a blade $A$, this operator defines what we will term the Hodge dual, given by:
\begin{equation}
	\star A = (\bm{\mathtt{A}} \widetilde{\bm{\mathtt{A}}})I,\nonumber
\end{equation}
where $\bm{\mathtt{A}}$ denotes the Euclidean component of $A$, i.e., the multivector obtained by setting the coefficient of the basis element $e$ to zero, and where $\widetilde{\bm{\mathtt{A}}}$ represents the reverse of $\bm{\mathtt{A}}$. By the linearity of the Hodge star operator, the Hodge dual of an arbitrary multivector is computed by decomposing the multivector into a sum of blades and applying the above computation to each blade \cite{Dor22}.

\subsection{Conformal geometric algebra}\label{CGA}
The conformal model extends the $n$-dimensional Euclidean space $\mathbb{R}^{n}$ by adding two extra basis vectors, $e$ and $\overline{e}$, with the property:
\begin{equation}
	e^{2}=1,\quad\overline{e}^{2}=-1.
\end{equation} 
These two extra vectors allow the definition of two null vectors:
\begin{equation}
	n_0=\dfrac{1}{2}\left(\overline{e}+e\right),\quad n_\infty=\overline{e}-e,
\end{equation}
where $n_\infty$ is associated with the point at infinity and $n_0$ with the origin. 

Thus, the conformal geometric algebra (CGA) of $\mathbb{R}^{n}$ is denoted by $\mathcal{G}_{n+1,1}$ and can be seen as the geometric algebra of $\mathbb{R}^{n+1,1}$. Now, the two null vectors $n_\infty$ and $n_0$ allow an intuitive description of translations, that are formulated as rotors in the same way as in PGA:
\begin{equation}\label{translatorCGA}
	T_{v} = 1 - \dfrac{v\wedge n_\infty}{2},
\end{equation}
where the translation is performed in the direction of $v\in\mathbb{R}^{n}$.

One of the most important advantages of conformal geometric algebra is that it provides a homogeneous model for the $n$-dimensional Euclidean space. In particular, every point $x\in\mathbb{R}^{n}$ is associated with a null vector of $\mathcal{G}_{n+1,1}$ (including the origin and the point at the infinity). This is done via the Hestenes' embedding:
\begin{equation}\label{Hestenes}
	\bm{x} = H(x) = \dfrac{1}{2}x^2n_\infty+n_0 +x,
\end{equation}
where $\bm{x}$ is said to be the null vector representation of $x$. The inverse of the Hestenes' embedding is just the projection operator onto $\mathbb{R}^n$. 

Another key feature of CGA is that it encodes geometric entities as elements of the algebra. In particular, the geometric entities of the conformal model of $\mathbb{R}^3$, $\mathcal{G}_{4,1}$, are considered, including points, lines, planes, circles and spheres. If $O$ denotes a geometric entity, then $\bm{o}$ is said to be the outer representation of $O$ if for every point $x\in O$, $\bm{x}\wedge\bm{o}= 0$. Taking the dual of the outer representation $\bm{o}$ of a geometric entity $O$\footnote{Here, since the pseudoscalar is constructed with $e$ and $\overline{e}$ instead of $n_0$ and $n_\infty$, there is no problem with taking the dual and no special pseudoscalar is needed.}, the inner representation $\bm{o}^\ast$ is obtained. The inner representation satisfies $\bm{x}\cdot\bm{o}^\ast=0$ for a null vector representation of $x\in O$.

Now, for two different geometric objects $O_1$ and $O_2$ with outer (inner) representations $\bm{o}_1$ and $\bm{o}_2$ ($\bm{o}_1^\ast$ and $\bm{o}_2^\ast$), their intersection, denoted by $\bm{o}_1\vee \bm{o}_2$, is the multivector:
\begin{equation}\label{intersection}
	\bm{o}_1\vee \bm{o}_2 = (\bm{o}_1^\ast\wedge \bm{o}_2^\ast)^\ast.
\end{equation}
Analogously, if the outer representations of $O_1$ and $O_2$ have the same grade, the angle defined by them is computed as follows:
\begin{equation}\label{angle}
	\angle(O_1,O_2) = \cos^{-1}\left(\dfrac{\bm{o}_1\cdot \bm{o}_2}{\sqrt{\bm{o}_1\widetilde{\bm{o}_1}}\sqrt{\bm{o}_2\widetilde{\bm{o}_2}}}\right).
\end{equation}
To describe the different geometric entities, let $p_1,p_2,p_3,p_4\in\mathbb{R}^3$ be four different points with null vector representation $\bm{p}_1,\bm{p}_2,\bm{p}_3,\bm{p}_4\in\mathcal{G}_{4,1}$. Then:
\begin{itemize}
	\item $\bm{b} = \bm{p}_1\wedge \bm{p}_2$ is a bivector and the outer representation of the pair of points $p_1$ and $p_2$.
	\item $\bm{\ell} = \bm{p}_1\wedge \bm{p}_2\wedge e_\infty$ is a trivector and the outer representation of the line passing through the points $p_1$ and $p_2$. Its inner representation is the bivector $\bm{\ell}^\ast= ve_{123} - (p_1\wedge v)e_{123}e_\infty$, where $v = p_1-p_2$ is its direction vector.
	\item $\bm{c} = \bm{p}_1\wedge \bm{p}_2\wedge \bm{p}_3$ is a trivector and the outer representation of a circle passing through the points $p_1, p_2$ and $p_3$. Its inner representation is the bivector $\bm{c}^\ast = \bm{\pi}^\ast\wedge \bm{s}^\ast$, where $\bm{\pi}^\ast$ and $\bm{s}^\ast$ are the inner representations of the plane and sphere whose intersection defines the circle.
	\item $\bm{\pi} = \bm{p}_1\wedge \bm{p}_2\wedge \bm{p}_3\wedge e_\infty$ is a 4-vector and the outer representation of a plane passing through the points $p_1, p_2$ and $p_3$. Its inner representation is the vector $\bm{\pi}^\ast = n + \delta e_\infty$, where $n$ denotes the unit vector normal to the plane and $\delta$, its orthogonal distance to the origin.
	\item $\bm{s} = \bm{p}_1\wedge \bm{p}_2\wedge \bm{p}_3\wedge \bm{p}_4$ is a 4-vector and the outer representation of a sphere passing through the points $p_1, p_2,p_3$ and $p_4$. Its inner representation is the vector $\bm{s}^\ast = \bm{z} - \frac{1}{2}r^2e_\infty$, where $\bm{z}$ is the null vector representation of the center of the sphere and $r$, its radius.
\end{itemize}

\section{State of the art}\label{s:soa}

There is a wider offering of computational libraries and tools for computing with GA. Probably the first documented library is CLICAL, a stand-alone calculator-like program running under MS-DOS~\cite{Lou87}. From this pioneer approach, many complementary implementations have appeared. In the following, some of them are reviewed with the objective of positioning $\mathtt{SUGAR}$ in the map, which can be almost completely drawn using the geometric algebra explorer website\footnote{\url{https://ga-explorer.netlify.app}} that shares a long list of GA-inspired software.

Looking at GA implementations for programming languages, different options exist, such as $\mathtt{Versor}$~\cite{Ver11}, a C++ library for geometric algebra, or $\mathtt{Ganja.js}$ \cite{GANJA}, a geometric algebra code generator for javascript capable of generating geometric algebras and subalgebras of any signature. $\mathtt{Ganja.js}$ also implements operator overloading and algebraic constraints. $\mathtt{GAlgebra}$ \cite{galgebra} is an implementation of a geometric algebra module in python that utilizes the sympy symbolic algebra library. However, it does not handle projective or conformal geometric algebras and is not maintained since 2019. Additionally, $\mathtt{GMac}$~\cite{Eid16}, short form for "Geometric Macro", is a sophisticated .NET based code generation software system that allows implementing geometric models and algorithms based on GA in arbitrary target programming languages. These implementations, in general, target programming languages that are not popular among engineers or mathematicians, that prefer for example VBA/VBS (scripting languages that stem from the Visual Basic programming) like MS Excel \footnote{\url{https://office.microsoft.com/excel}}, Labview \footnote{\url{https://www.ni.com/es/shop/labview.html}}, or the already mentioned Matlab.

In other cases, GA is implemented using a specialized package, either symbolic or numeric, within a larger mathematical software system. For instance, $\mathtt{Clifford}$~\cite{Pro16} is a lightweight package for performing geometric algebra calculations in Maxima \footnote{\url{https://maxima.sourceforge.io/index.html}} -- a computer algebra system for the manipulation of symbolic and numerical expressions.       $\mathtt{CLIFFORD}$~\cite{Abl05} is a package for Clifford and Grassmann algebras computations within Maple\footnote{\url{https://https://www.maplesoft.com}} -- a symbolic and numeric computing environment, and $\mathtt{CGAlgebra}$~\cite{Ara22} is a package for 5D conformal geometric algebra in Mathematica \footnote{\url{https://www.wolfram.com/mathematica}} -- a platform for technical computing that has been the basis for the development of the WolframAlpha answer engine.

Regarding Matlab, GABLE, which stands for Geometric Algebra Learning Environment, is the first geometric algebra package mentioned in the literature \cite{Mann99}. However, its main limitation is that it only handles geometric algebras up to $p+q=3$. Additionally, it can only process single multivectors, i.e., it cannot handle arrays or matrices of multivectors. An additional implementation of a geometric algebra package covers basic algebraic operations, allowing symbolic manipulations, but it supports only up $p+q=4$ \cite{Ant14}. The limitation rises from the implementation of the algebraic operations, which are explicitly coded for each dimension. A qualitative  progress is found in the Clifford Multivector Toolbox \cite{San16} (from the same authors of the Quaternion Toolbox for Matlab, QTM \cite{qtfm}), which has been designed to extend Matlab in a natural Matlab-like manner to handle  arrays (including, but not limited to, vectors and matrices) with elements which are Clifford or geometric multivectors in an arbitrarily chosen geometric algebra. The main limitation relies in the lack of support for symbolic computations. In addition, its treatment of PGA and CGA does not allow a full geometric interpretation of the manipulations of the special elements from those algebras, which may discourage mathematicians and engineers from using it. To fill this gap, $\mathtt{SUGAR}$ offers symbolic and numerical computations, permits operating with arrays or matrices of multivectors, and many of the functions that implements are overloading of existing Matlab functions, which eases its applicability, making it user-friendly. In addition, PGA and CGA has been designed so that their use is done in the same way (i.e., using the same formulas and expressions) as in the literature.

Therefore, the reasons behind the creation of $\mathtt{SUGAR}$ are, as stated before, the need for a software able to deal with the computations (sometimes symbolic) that arise in the different research fields of the authors and, by extension, research fields within mathematics, applied mathematics, and engineering whose open problems can be addressed using geometric algebra. In particular, the authors have developed strategies based on geometric and conformal geometric algebra in applications such as molecular geometry, space-time physics, modeling, analysis and control of three-phase electronic circuits, and robotics \cite{Lav18,Vel23,Zap21, Isi22}.

\section{Overview of the $\mathtt{SUGAR}$ toolbox}\label{s:des}
$\mathtt{SUGAR}$ has been developed by the authors since June 2022 and has been publicly released in December 2023 at \url{https://github.com/distributed-control-systems/SUGAR}. The current state of implementation is:
\begin{itemize}
	\item the underlying structure of the toolbox is complete (initialization and computation with GA, PGA and CGA of any signature);
	\item creation of the conformal model of any pseudo-Euclidean space $\mathbb{R}^{p,q}$ is available (not limited to the usual $\mathbb{R}^2$ or $\mathbb{R}^3$);
	\item the definition of multivectors by the user can be done intuitively or using a specific constructor, and grade extraction, and involution operator (e.g., conjugate, reverse) are implemented;
	\item several basic arithmetic operations and functions for multivectors have been overloaded;
	\item arithmetic functions for multivectors are implemented, including the full geometric product of two multivectors;
	\item non-trivial trigonometric and exponential functions such as $\sin$, $\cos$, $\log$, $\exp$ and the like are also available;
	\item the possibility of coding new user-defined functions on multivectors is available;
	\item all functions are available for any dimension (as long as the user has enough computer power), numerically or symbolically;
	\item the matrix-handling of Matlab has been extended to work with multivectors, and therefore standard computations of arrays and matrix of multivectors are available.
\end{itemize}

Some of the previous features, such as the implementation of non-trivial functions, as well as their application to high-dimensional algebras, are facilitated by the internal representation that multivectors have in $\mathtt{SUGAR}$, which relies on matrices. If the basis elements of $\mathcal{G}_{p,q,r}$ are numbered in order of construction, i.e., $E_0 = e_0$, $E_1 = e_1,\dots E_{p+q+r} = e_{12\dots n}$, then a given multivector $A\in\mathcal{G}_{p,q,r}$ can be expressed as $A = \sum_i A_iE_i$ with $0\leq i\leq p+q+r$. This allows to easily represent $A$ by a $2^{p+q+r}\times 2^{p+q+r}$ matrix $M_A$\cite{Roe21}:
\begin{equation}\label{eq:matrix_representation}
	M_A= \begin{bmatrix}
		\vrule & \vrule & \vrule & \dots & \vrule\\
		[AE_0] & [AE_1] & [AE_2] & \dots & [AE_{p+q+r}]\\
		\vrule & \vrule & \vrule & \dots & \vrule
	\end{bmatrix}
\end{equation}
where $[\cdot]$ denotes the component extraction operator, i.e., $[AE_i]$ is an array with the coefficients of multivector $AE_i$ in ascending order. This, in turn, establishes a faithful representation and, hence, an isomorphism, between any geometric algebra $\mathcal{G}_{p,q,r}$ and the algebra of matrices of order $2^{p+q+r}$, $\mathcal{M}_{2^{p+q+r}}$.

\begin{rem}
	This representation is not unique. The reader is referred to the work of Calvet \cite{Cal17}, where the analysis of faithful representations of Clifford algebras is used to establish the minimum and maximum order of the matrix algebras for which a given geometric algebra can be isomorphic via a matrix representation. However, if matrices of order $2^{p+q+r}$ are considered, this representation is clearly unique.
\end{rem}

The $\mathtt{SUGAR}$ toolbox  has two basic utilities for creating a geometric algebra (GA), a projective geometric algebra (PGA), or a conformal geometric algebra (CGA). In addition,  it has a multivector class, named $\mathtt{MV}$, whose methods, functions and operators permit all kind of manipulations and computations. The installation requires having the folder named $\mathtt{SUGAR}$ that contains the $\mathtt{@MV}$, $\mathtt{Examples}$ and $\mathtt{Utility}$ subfolders included in the $\mathtt{Matlab}$ path.

\subsection{Creating an algebra}
The creation of a GA is done through the $\mathtt{GA(signature,options)}$ function. This function expects as a first parameter the number of basis vectors that square to $+1$ (positive square), $-1$ (negative square), and $0$ (null square), i.e., the {\it signature} $\mathtt{[p, q, r]}$.  Optionally, it admits a second optional parameter, $\mathtt{"verbose"}$, that tells you what has been done. The function creates the specified GA of the pseudo-Euclidean vector space $\mathbb{R}^{p+q+r}$, whose basis elements are $e_0, e_1, \hdots, e_p, e_{p+1}, \hdots e_{p+q}, e_{p+q+1}, \hdots, e_{p+q+r}$, and all their products  with indices in strict ascending order. In turn, the verbose states that all $2^{p+q+r}$ basis elements have been created, it lists them and also states the different grades that can be found in a general multivector generated by the given GA. For instance, the creation of $\mathcal{G}_{2,0,0}$ is:
\begin{colorverbatim}
	>> GA([2,0,0],"verbose")
	Declaring e0 as syntatic sugar, e0=1
	Declaring e1 such that e1·e1=1
	Declaring e2 such that e2·e2=1
	Declaring e12 such that e12·e12=-1
	
	Declaring G0 for grade slicing as (1)*e0 
	Declaring G1 for grade slicing as (1)*e1+(1)*e2 
	Declaring G2 for grade slicing as (1)*e12 
\end{colorverbatim}
\noindent The creation of a particular instance of a PGA is done using the same function, with the only difference that the signature must be $[n,0,1]$ for $n\in\mathbb{N}$, i.e., PGA consists of adding a null vector to an Euclidean space $\mathbb{R}^n$ and computing its associated geometric algebra. 

\begin{figure}[h]
	\centering
	\includegraphics[scale=0.45]{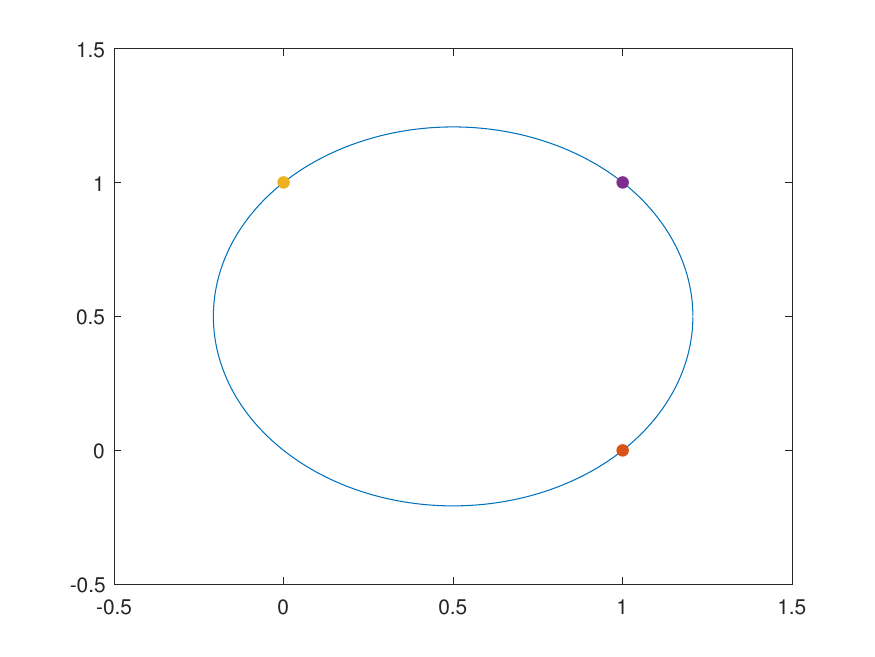}
	\caption{Plot of the circle defined by the Euclidean points $e_1$ (orange), $e_2$ (yellow), and $e_1+e_2$ (violet).\label{plot_circle}}
\end{figure}

The creation of a CGA is done via a dedicated function, the $\mathtt{CGA(signature,options)}$ function. This function expects as a first parameter either a scalar, usually $2$ or $3$, that refers to the $\mathbb{R}^2$ o $\mathbb{R}^3$  vector spaces whose conformal model is going to be constructed, or, as before, the signature, $\mathtt{[p, q, r]}$. In fact, this allows to conformalize any pseudo-Euclidean vector space. Optionally, it admits a second parameter, $\mathtt{"verbose"}$, that indicates what has been done. For instance, it explains that all the $2^{p+1+q+1+r}$ basis elements have been created and states the different grades that can be found in a general multivector of that particular algebra. In addition, it lists the $\mathtt{push}$ and $\mathtt{pull}$ operators, which can be used to map vectors form GA to null vectors of CGA (Hestenes' embedding) and vice-versa (inverse Hestenes' embedding), respectively. The creation of the conformal algebra from $\mathbb{R}^2$ is:
\begin{colorverbatim}
	>> CGA([2,0,0],"verbose")
	
	---- CGA BASIS -----
	Declaring e0 as syntactic sugar, e0=1
	Declaring n0 such that n0·n0=0
	Declaring e1 such that e1·e1=1
	Declaring e2 such that e2·e2=1
	Declaring ni such that ni·ni=0
	Declaring n0e1 such that n0e1·n0e1=0
	Declaring n0e2 such that n0e2·n0e2=0
	Declaring n0ni such that n0ni·n0ni=1
	Declaring e12 such that e12·e12=-1
	Declaring e1ni such that e1ni·e1ni=0
	Declaring e2ni such that e2ni·e2ni=0
	Declaring n0e12 such that n0e12·n0e12=0
	Declaring n0e1ni such that n0e1ni·n0e1ni=1
	Declaring n0e2ni such that n0e2ni·n0e2ni=1
	Declaring e12ni such that e12ni·e12ni=0
	Declaring n0e12ni such that n0e12ni·n0e12ni=-1
	
	Declaring G0 for grade slicing as (1)*e0 
	Declaring G1 for grade slicing as (1)*n0+(1)*e1+(1)*e2+(1)ni 
	Declaring G2 for grade slicing as (1)*n0e1+(1)*n0e2+(1)*n0ni+(1)*e12+(1)*e1ni+(1)*e2ni 
	Declaring G3 for grade slicing as (1)*n0e12+(1)*n0e1ni+(1)*n0e2ni+(1)*e12ni 
	Declaring G4 for grade slicing as (1)*n0e12ni 
	
	push and pull operations are now available 
\end{colorverbatim}
\noindent Once a CGA has been created, the $\mathtt{push}$  and $\mathtt{pull}$  functions can be used to obtain a multivector representation in either the original GA or its conformal model. For example, after creating $\mathtt{CGA([2,0,0],"verbose")}$, a vector $\mathtt{p}$ in GA can be represented as a null vector of CGA, $\mathtt{pc}$, by using the $\mathtt{push}$ operator, and vice-versa. The code and results for a particular vector $\mathtt{p}$ are as follows:
\begin{colorverbatim}
	>> p=e1+e2
	p = 
	( 1 )*e1+( 1 )*e2
	>> pc=push(p)
	pc = 
	( 1 )*n0+( 1 )*e1+( 1 )*e2+( 1 )*ni
	>> pbis=pull(pc)
	pbis = 
	( 1 )*e1+( 1 )*e2
\end{colorverbatim}
\noindent Finally, the CGA module includes a visualization tool, accessible via the command \verb|plot()|, that displays CGA elements representing various geometric entities that can be encoded in CGA. This feature is useful for validating or verifying steps in the development of geometric solutions. The code, results, and plot (Fig. \ref{plot_circle}) for a circle $S$ passing through the Euclidean points $e_1,e_2$, and $e_1+e_2$ are as follows:
\begin{colorverbatim}
	>> CGA([2 0 0])
	>> B=e1;C=e2;D=e1+e2;                           % Definition of three Euclidean points
	>> B_CGA=push(B);C_CGA=push(C);D_CGA=push(D);   % Associated null vectors
	>> S = B_CGA.^C_CGA.^D_CGA                      % Definition of a circle passing through the
	S =                                             % three points
	( -1 )*n0e12 +( -0.5 )*n0e1ni +( 0.5 )*n0e2ni 
	>> S.plot(),xlim([-1,2]),ylim([-1,2])           % Plot of the computed circle
\end{colorverbatim}

\subsection{Creating multivectors}

Multivectors are defined using the $\mathtt{@MV}$ class. As stated before, every multivector of a GA (and of PGA and CGA) can be expressed as a linear combination of the basis elements of the algebra. Therefore, they can be defined in $\mathtt{SUGAR}$ using this rule. For example, within the GA $\mathcal{G}_{3,0,0}$, a multivector $A$ can be defined as
\begin{colorverbatim}
	>> GA([3,0,0]); 
	>> A=e1+5*e2+4*e12-7*e123
	A = 
	(1)*e1+(5)*e2+(4)*e12+(-7)*e123
\end{colorverbatim}
\noindent Alternatively, a multivector can also be defined using the constructive method $\mathtt{MV(val, signature, representation)}$, where $\mathtt{val}$ is an array containing all coefficients, $\mathtt{signature}$ specifies the $\mathtt{[p, q, r]}$ signature, and $\mathtt{representation}$ serves to specify whether the algebra is just a GA or PGA, with an empty parameter value, or a CGA, with the string $\mathtt{"CGA"}$ as a parameter value. For example, within the same GA, a multivector $B$ can be defined as:
\begin{colorverbatim}
	>> B=MV([3 8 0 -5 0 4 -2 -1],[3,0,0])
	B = 
	(3)*e0+(8)*e1+(-5)*e3+(4)*e13+(-2)*e23+(-1)*e123
\end{colorverbatim}
\noindent The second method differs from the first primarily because it does not require initializing the algebra. In the first method, SUGAR creates a variable for each basis element, resulting in a total of $2^{p+q+r}$ variables stored in the MATLAB workspace. This process, which we refer to as initializing the algebra, can generate a large number of variables in high-dimensional algebras, potentially leading to memory overhead. In contrast, the second method avoids declaring and storing these numerous variables, which can be useful for high-dimensional algebras. In addition, this second method throws an exception error if the length of $\mathtt{val}$ does not coincide with the value $2^{p+q+r}$, where $p$, $q$, and $r$ represent the elements of the signature $\mathtt{[p, q, r]}$. Furthermore, both methods for multivector definition allow for the use of symbolic coefficients, as shown in the following example:
\begin{colorverbatim}
	>> syms a1 a2 a3 a4 a5 a6 a7 a8; 
	>> C=MV([a1 a2 a3 a4 a5 a6 a7 a8],[3,0,0])
	C = 
	(a1)*e0+(a2)*e1+(a3)*e2+(a4)*e3+(a5)*e12+(a6)*e13+(a7)*e23+(a8)*e123
\end{colorverbatim}
\noindent For CGA, the definition of a multivector follows the same rules but using either the basis elements:
\begin{colorverbatim}
	>> syms a b c d real; 
	>> p=a*n0+b*e1+c*e2+d*ni
	p = 
	(a)*n0+(b)*e1+(c)*e2+(d)*ni
\end{colorverbatim}
\noindent or the $\mathtt{MV}$ constructor with the additional $\mathtt{"CGA"}$ parameter:
\begin{colorverbatim}
	>> MV([0 a b c d zeros(1,11)],[3,1,0],"CGA")
	ans = 
	(a)*n0+(b)*e1+(c)*e2+(d)*ni
\end{colorverbatim}
\noindent The properties associated with a multivector are the $\mathtt{signature}$ of the algebra where it belongs, its $\mathtt{coefficients}$, its $\mathtt{matrix}$ representation and the $\mathtt{BasisNames}$. These properties can be obtained using the command $\mathtt{multivector.property}$. For instance, to obtain the names of the basis elements of the previously computed multivector $C$ or the matrix representation of the previously computed multivector $B$, the introduced command can be used as:
\begin{colorverbatim}
	>> C.BasisNames
	ans =
	1×8 cell array
	{["e0"]}  {["e1"]}  {["e2"]}  {["e3"]}  {["e12"]}  {["e13"]}  {["e23"]}  {["e123"]}
	
	>> B.matrix
	
	ans =
	
	3     8     0    -5     0    -4     2     1
	8     3     0     4     0     5     1     2
	0     0     3    -2     8    -1     5     4
	-5    -4     2     3     1     8     0     0
	0     0     8    -1     3    -2    -4    -5
	4     5     1     8     2     3     0     0
	-2    -1     5     0     4     0     3     8
	-1    -2    -4     0    -5     0     8     3
\end{colorverbatim}
\noindent In addition, $\mathtt{SUGAR}$ allows a natural slicing, i.e., selection of coefficients, for any defined multivector. The indexing is vector-based and uses the standard Matlab notation. It is based in the numbering of the basis elements defined before equation \eqref{eq:matrix_representation}. In addition, the coefficient of a basis element can also be extracted using, instead of an index or set of indices, a basis name. The following code is an example of the extraction of a given multivector coefficients using the two described ways:

\begin{colorverbatim}
	>> syms x y z t real
	>> GA([2,0,0]) 
	>> A=x*e0+y*e1+z*e2+t*e12 
	A = 
	(x)*e0+(y)*e1+(z)*e2+(t)*e12
	>> A(2)
	ans =
	y
	>> A(2:4)
	ans =
	[y, z, t]
	>> A(e0)
	ans=
	x
\end{colorverbatim}
\noindent In addition, curly brackets can be used to extract sub-multivectors attending to their position in the original multivector:
\begin{colorverbatim}
	>> A{2}
	ans = 
	(y)e1
	>> A{2:4}
	ans = 
	(y)*e1+(z)*e2+(t)*e12
\end{colorverbatim}
\noindent Finally, normal and curly brackets can be used to extract $k$-vectors, for different values of $k$, from the original multivector:
\begin{colorverbatim}
	>> A(G1)
	ans =
	[y, z]
	>> A{G1}
	ans = 
	(y)*e1+(z)*e2
\end{colorverbatim}

\subsection{Basic operations with multivectors}

The plus sign $+$ and the minus sign $-$ represent the operation of addition and subtraction of two multivectors, which results in their sum and difference, respectively. For example, the addition of the previously defined $A$ and $B$ multivectors is given by:
\begin{colorverbatim}
	>> A+B
	ans = 
	(3)*e0+(9)*e1+(5)*e2+(-5)*e3+(4)*e12+(4)*e13+(-2)*e23+(-8)*e123
\end{colorverbatim}
\noindent The key operation over multivectors is the geometric product, which is denoted in $\mathtt{SUGAR}$ by $\mathtt{*}$. For instance, $\mathcal{G}_{2,0,0}^+=\text{span}\{e_0,e_{12}\}$ is isomorphic to the complex numbers, and therefore, the geometric product of two multivectors $C1,C2\in\mathcal{G}_{2,0,0}^+$ is the same as the product between two complex numbers with the same coefficients as $C1$ and $C2$:
\begin{colorverbatim}
	>> GA([2,0,0]); 
	>> C1=1+2*e12;
	>> C2=5-1*e12;
	>> C3=C1*C2
	C3 = 
	(7)*e0+(9)*e12
	>> z1 = 1 + 2i;
	>> z2 = 5 - i;
	>> z3 = z1*z2
	z3 = 7 + 9i
\end{colorverbatim}
\noindent Analogously, the inner and outer products are done using the overloaded operators $.*$ and $.^\wedge$. For example, the inner and outer products of two multivectors of $\mathcal{G}_{3,0,0}$ can be computed as follows:
\begin{colorverbatim}
	>> GA([2 0 0])
	>> D1=2*e0+3*e1+2*e2+4*e12
	D1 = 
	( 2 )*e0+( 3 )*e1+( 2 )*e2+( 4 )*e12
	>> D2=1*e0-2*e1+1*e2+3*e12
	D2 = 
	( 1 )*e0+( -2 )*e1+( 1 )*e2+( 3 )*e12
	>> D1.*D2
	ans = 
	( -14 )*e0+( -3 )*e1+( 21 )*e2+( 10 )*e12
	>> D1.^D2
	ans = 
	( 2 )*e0+( -1 )*e1+( 4 )*e2+( 17 )*e12
\end{colorverbatim}
\noindent Since the division between two multivectors is essentially the product of the dividend by the inverse of the divisor, and given that the geometric product is not commutative, the operator $/$ cannot be used (as it does not define the precedence of the operands). Hence, division should be performed by applying the geometric product of one multivector (the dividend) by the inverse of the other multivector (the divisor). In addition, since this inverse involves computing the power of a multivector with an exponent of $-1$, powers can be generalized to any integer. This operation is achieved by overloading the $^\wedge$ operator. For example, the following code illustrates the division of the multivector $A$ by the multivector $B$:
\begin{colorverbatim}
	>> A = sym(1/5)+sym(2/5)*e1+sym(2/5)*e2+sym(4/5)*e12;
	>> B = sym(1/2)-sym(1/2)*e1+sym(1/2)*e2+sym(1/2)*e12;
	>> Binv=B^-1
	ans = 
	(1/2)*e0+(1/2)*e1+(-1/2)*e2+(-1/2)*e12
	A*Binv
	ans = 
	(1/2)*e0+(1/2)*e1+(7/10)*e2+(-1/10)*e12
\end{colorverbatim}
\noindent Now, it is easy to check that a multivector multiplied by its inverse is the identity:
\begin{colorverbatim}
	>> B*Binv
	ans = 
	(1)*e0
\end{colorverbatim}
\noindent Finally, recall that $\mathtt{SUGAR}$ deals with GA of high dimension. For example, the computation of the inverse of a multivector  $C\in\mathcal{G}_{7,0,0}$ can be done easily as shown in the following piece of code:
\begin{colorverbatim}
	>> C=MV.rand([7 0 0])
	C = 
	(0.81472)*e0+(0.90579)*e1+(0.12699)*e2+(0.91338)*e3+(0.63236)*e4+(0.09754)*e5+(0.2785)*e6... 
	+(0.54688)*e7+(0.95751)*e12+(0.96489)*e13+(0.15761)*e14+(0.97059)*e15+(0.95717)*e16 ...
	+(0.48538)*e17+(0.80028)*e23+(0.14189)*e24+(0.42176)*e25+(0.91574)*e26+(0.79221)*e27 ...
	+(0.95949)*e34+(0.65574)*e35+(0.035712)*e36+(0.84913)*e37+(0.93399)*e45+(0.67874)*e46 ...
	+(0.75774)*e47+(0.74313)*e56+(0.39223)*e57+(0.65548)*e67+(0.17119)*e123+(0.70605)*e124 ...
	+(0.031833)*e125+(0.27692)*e126+(0.046171)*e127+(0.097132)*e134+(0.82346)*e135 ...
	+(0.69483)*e136+(0.3171)*e137+(0.95022)*e145+(0.034446)*e146+(0.43874)*e147 ...
	+(0.38156)*e156+(0.76552)*e157+(0.7952)*e167+(0.18687)*e234+(0.48976)*e235+(0.44559)*e236...
	+(0.64631)*e237+(0.70936)*e245+(0.75469)*e246+(0.27603)*e247+(0.6797)*e256+(0.6551)*e257...
	+(0.16261)*e267+(0.119)*e345+(0.49836)*e346+(0.95974)*e347+(0.34039)*e356+(0.58527)*e357...
	+(0.22381)*e367+(0.75127)*e456+(0.2551)*e457+(0.50596)*e467+(0.69908)*e567+(0.8909)*e1234...
	+(0.95929)*e1235+(0.54722)*e1236+(0.13862)*e1237+(0.14929)*e1245+(0.25751)*e1246...
	+(0.84072)*e1247+(0.25428)*e1256+(0.81428)*e1257+(0.24352)*e1267+(0.92926)*e1345...
	+(0.34998)*e1346+(0.1966)*e1347+(0.25108)*e1356+(0.61604)*e1357+(0.47329)*e1367...
	+(0.35166)*e1456+(0.83083)*e1457+(0.58526)*e1467+(0.54972)*e1567+(0.91719)*e2345...
	+(0.28584)*e2346+(0.7572)*e2347+(0.75373)*e2356+(0.38045)*e2357+(0.56782)*e2367...
	+(0.075854)*e2456+(0.05395)*e2457+(0.5308)*e2467+(0.77917)*e2567+(0.93401)*e3456...
	+(0.12991)*e3457+(0.56882)*e3467+(0.46939)*e3567+(0.011902)*e4567+(0.33712)*e12345...
	+(0.16218)*e12346+(0.79428)*e12347+(0.31122)*e12356+(0.52853)*e12357+(0.16565)*e12367...
	+(0.60198)*e12456+(0.26297)*e12457+(0.65408)*e12467+(0.68921)*e12567+(0.74815)*e13456...
	+(0.45054)*e13457+(0.083821)*e13467+(0.22898)*e13567+(0.91334 )*e14567...
	+(0.15238)*e23456+(0.82582)*e23457+(0.53834)*e23467+(0.99613)*e23567+(0.078176)*e24567...
	+(0.44268)*e34567+(0.10665)*e123456+(0.9619)*e123457+(0.0046342)*e123467...
	+(0.77491)*e123567+(0.8173)*e124567+(0.86869)*e134567+(0.084436)*e234567...
	+(0.39978)*e1234567
	>> Cinv=C^-1;
	>> clean(C*Cinv)
	ans = 
	( 1 )*e0
\end{colorverbatim}
\noindent where the Matlab function $\mathtt{clean}$ has been used to remove terms that are zero to machine precision ($\mathtt{e-16}$) when doing numerical computations.

In the following example, we measure the execution time required to initialize geometric algebras with signatures $(6,0,0),(7,0,0),(8,0,0),(9,0,0)$, and $(9,1,0)$. In addition, we compute the execution time for declaring a random multivector in each of these algebras and inverting it. As expected, the initialization time increases with the algebra dimension; however, once initialized, computations remain efficient. Notably, for dimensions up to $p+q+r=8$, initialization remains relatively fast\footnote{ These times can vary depending on the computer on which the script is executed. However, experiments carried out by the authors with different processors show that for algebras up to $p+q+r=8$, there is barely any difference.}.
\begin{colorverbatim}
	tic
	GA([6 0 0])
	toc
	Elapsed time is 0.790655 seconds.
	tic
	a = MV.rand([6,0,0]);
	b = a^-1;
	toc
	Elapsed time is 0.002825 seconds.
	
	tic
	GA([7 0 0])
	toc
	Elapsed time is 1.108590 seconds.
	tic
	a = MV.rand([7,0,0]);
	b = a^-1;
	toc
	Elapsed time is 0.005595 seconds.
	
	tic
	GA([8 0 0])
	toc
	Elapsed time is 4.122144 seconds.
	tic
	a = MV.rand([8,0,0]);
	b = a^-1;
	toc
	Elapsed time is 0.004904 seconds.
	
	tic
	GA([9 0 0])
	toc
	Elapsed time is 18.076994 seconds.
	tic
	a = MV.rand([9,0,0]);
	b = a^-1;
	toc
	Elapsed time is 0.012538 seconds.
	
	tic
	GA([9 1 0])
	toc
	Elapsed time is 95.988849 seconds.
	tic
	a = MV.rand([9,1,0]);
	b = a^-1;
	toc
	Elapsed time is 0.180647 seconds.
\end{colorverbatim}
\begin{rem}
	The computation of symbolic multivector inverses in geometric algebras satisfying $p+q+r>5$ with $\mathtt{SUGAR}$ can be employed to derive closed-form formulas for computing such inverses. Existing formulas found in the literature only extend up to $p+q+r=5$ \cite{Hitzer_inv}. The authors are currently preparing an article further exploring this idea.
\end{rem}
\subsection{Functions of multivectors}
Basic Matlab functions have been overloaded to allow operating with numeric or symbolic multivectors. Some of these functions include $\sin$, $\mathrm{asin}$, $\sinh$, $\mathrm{asinh}$, $\cos$, $\mathrm{acos}$, $\cosh$, $\mathrm{acosh}$,  $\tan$, $\mathrm{atan}$, $\tanh$, $\mathrm{atanh}$, $\cot$, $\mathrm{acot}$, $\coth$, $\mathrm{acoth}$, $\csc$, $\mathrm{acsc}$, $\mathrm{csch}$, $\mathrm{acsch}$, $\sec$, $\mathrm{asec}$, $\mathrm{sech}$, $\mathrm{asech}$, $\log$, and $\exp$. In the following code, the exponential of a randomly generated multivector $A$ is computed first, and then the logarithm of the previous result is computed, leading again to $A$:
\begin{colorverbatim}
	>> A=MV.rand([2 0 0])
	A = 
	(0.81472)*e0+(0.90579)*e1+(0.12699)*e2+(0.91338)*e12
	>> exp(A)
	ans = 
	(2.2612)*e0+(2.0466)*e1+(0.28692)*e2+(2.0637)*e12
	>> log(ans)
	ans = 
	(0.81472)*e0+(0.90579)*e1+(0.12699)*e2+(0.91338)*e12
\end{colorverbatim}
\noindent Analogously, the exponential of a symbolic multivector is computed as follows:
\begin{colorverbatim}
	>> syms a [1 4] 
	>> A=MV(a,[2 0 0])
	A = 
	(a1)*e0+(a2)*e1+(a3)*e2+(a4)*e12
	>> exp(A)
	ans = 
	(exp(a1+(a2^2+a3^2-a4^2)^(1/2))/2+exp(a1-(a2^2+a3^2-a4^2)^(1/2))/2)*e0...
	+((a2*exp(a1+(a2^2+a3^2-a4^2)^(1/2))- ...
	a2*exp(a1-(a2^2+a3^2-a4^2)^(1/2)))/(2*(a2^2+a3^2-a4^2)^(1/2)))*e1...
	+((a3*exp(a1+(a2^2+a3^2-a4^2)^(1/2))- ...
	a3*exp(a1-(a2^2+a3^2-a4^2)^(1/2)))/(2*(a2^2+a3^2-a4^2)^(1/2)))*e2...
	+((a4*exp(a1+(a2^2+a3^2-a4^2)^(1/2))- ...
	a4*exp(a1-(a2^2+a3^2-a4^2)^(1/2)))/(2*(a2^2+a3^2-a4^2)^(1/2)))*e12
\end{colorverbatim}
\begin{rem}
	To date, there are no closed-form formulas for the exponential of general multivectors in geometric algebras with either $p+q+r\leq5$ or $p+q+r>5$. In particular, closed-form formulas have been developed for certain algebras with $p+q+r\leq5$ or for particular blades in algebras with $p+q+r\leq5$ \cite{exponentials}. As in the case of multivector inverses, the symbolic power of $\mathtt{SUGAR}$ can be used to compute closed-form formulas for the exponential of general multivectors in geometric algebras of any signature.
\end{rem}

\subsubsection{Matrices of multivectors}
By exploiting the Matlab-based array-inspired computations, $\mathtt{SUGAR}$ allows the creation of matrices where the components are multivectors of the same GA as well as performing operations with them, as demonstrated in the following example:

\begin{colorverbatim}
	>> GA([1,1,0])
	>> M=[e1 e1+e2; e2 e2-e1]
	M = 
	(1)*e1    (1)*e1+(1)*e2    
	(1)*e2    (-1)*e1+(1)*e2  
	>> M*M^-1
	ans = 
	(1)*e0    0     
	0     (1)*e0 
\end{colorverbatim}
\noindent Multivectors matrix operations can be done numerically or symbolically in the same way as with non-multivector matrices.

\subsection{Multivector methods}
This section presents Table~\ref{t:fun} which lists the main functions/methods/operations available for multivectors (and often for matrices of multivectors). They are not described in detail, but an intuitive description is provided. The following example illustrates the use of some of the functions listed in Table~\ref{t:fun}:

\begin{colorverbatim}
    >> GA([2,0,0]);
	>> logexp=@(y)y.apply(@(x)log(exp(x)));
	>> logexp(20*e1)
	ans = 
	( 20 )*e1
	>> p=e0+e1+e2+e12;
	>> grade(p,1)
	ans = 
	( 1 )*e1+( 1 )*e2
	>> p.dual
	ans = 
	( 1 )*e0+( 1 )*e1+( -1 )*e2+( 1 )*e12
	>> syms t real; p=t*e0+sin(4*t*e1); p.laplace
	ans = 
	( 1/s^2 )*e0+( 4/(s^2 + 16) )*e1
\end{colorverbatim}
\begin{table}[h!]  
	\caption{List of functions.}\label{t:fun}
	\begin{center}
		\begin{tabular}{ l | l }
			\hline	
			$\mathtt{abs}$ & returns a multivector whose coefficients are given in absolute value\\
			$\mathtt{apply}, \mathtt{\&}$ & applies a (user-defined) function to a multivector\\
			$\mathtt{and}$ & computes the "vee" or GA commutator product of two multivectors\\
			$\mathtt{clean}$ & remove those nasty-like terms ($e-16$) when doing numerical computations\\   
			$\mathtt{conj}, \mathtt{\text{'}}$ & computes the geometric conjugate of a mulivector \cite{Hitzer_inv}\\   
			$\mathtt{ctranspose}, \mathtt{\text{.'}}$ & computes the complex conjugate transpose of a matrix of multivectors\\   
			$\mathtt{det}$ & computes the determinant of a multivector matrix\\
			$\mathtt{dual}$ & computes the dual of a multivector \\
			$\mathtt{eq}, \mathtt{==}$ & determines if two multivectors are equal\\
			$\mathtt{expand}$ & (symbolically) expands the coefficients of the multivector\\
			$\mathtt{grade}$ & returns the $k$-vector of a multivector for a given $k$\\
			$\mathtt{ilaplace}$ & computes the Laplace inverse of the coefficients of a multivector\\
			$\mathtt{info}$ & recovers the information of a given geometric entity from the conformal model of $\mathbb{R}^2$ or $\mathbb{R}^3$\\
			$\mathtt{inv}, \mathtt{\wedge-1}$ & computes the inverse of a multivector or a matrix of multivectors\\
			$\mathtt{laplace}$ & computes the Laplace transform of the coefficients of a multivector \\
			$\mathtt{latex}$ & provides the latex expression of a multivector\\
			$\mathtt{length}$ & computes the magnitude/norm of a multivector\cite{Hes84}\\
			$\mathtt{maininvolution}$ & computes the main involution of a multivector\cite{Lav18}\\
			$\mathtt{normalize}$ & normalizes a multivector (i.e., it divides the multivector by its magnitude/norm)\\
			$\mathtt{not}$ & computes the reverse of a multivector\\
			$\mathtt{pinv}$ & computes the pseudo-inverse of a matrix of multivectors\\
			$\mathtt{plot}$ & plots a given geometric entity from the conformal model of $\mathbb{R}^2$ or $\mathbb{R}^3$\\
			$\mathtt{reverse}, \mathtt{\thicksim}$ & computes the reverse of a multivector \\
			$\mathtt{simplify}$ & simplifies the coefficients of a multivector\\
			$\mathtt{sqrt}$ & computes the square root of a multivector\\
			$\mathtt{str}$ & provides a text string that represents the multivector\\
			$\mathtt{vector}$ & returns an array with the coefficients of a multivector\\
			\hline  
		\end{tabular}
	\end{center}
\end{table}
\section{Application examples}\label{s:app}

This section provides various examples demonstrating the applicability of $\mathtt{SUGAR}$ across different domains, specifically in the realms of robotics and power electronics. 

The initial two application examples center around robot kinematics, where kinematic equations establish the relationship between the robot's joint variables and the position and orientation of its end-effector. Robot kinematics can be divided into forward and inverse kinematics. Forward kinematics is the problem consisting of calculating the position and orientation of the end-effector given the current joint positions or configuration of the robot, while inverse kinematics consists of the determination of the set of joint variables or robot's configurations given the position and orientation of the end-effector. Due to their geometric nature, these examples naturally align with the tools provided by projective and conformal geometric algebra, offering a comprehensive taste of the capabilities of $\mathtt{SUGAR}$.

An additional example complements this section, focusing on power electronics. It demonstrates SUGAR's capability to operate with matrices of multivectors, both symbolically and numerically.

\subsection{Forward kinematics of a 3R planar robot arm}\label{ex:FK}

\begin{figure}[h]
	\centering
	\includegraphics[width=6cm]{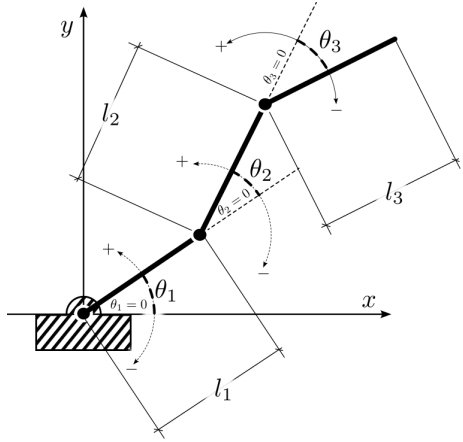}
	\caption{Schematic representation of a 3R planar robot.\label{fig:2Drobot}}
\end{figure}

The objective of this example is to illustrate the user-friendly capabilities of $\mathtt{SUGAR}$ in manipulating PGA elements (both numerical and symbolic) for solving a classical problem in robotics. Consider a planar robot arm as the one depicted in Fig.~\ref{fig:2Drobot}. This robot arm consists of three rigid links with lengths $\ell_1$, $\ell_2$, and $\ell_3$, and three planar revolute joints connected in series. Each joint connects two consecutive links, allowing the following link to rotate relative to the preceding link within the same plane. Each joint is defined by the counter-clockwise angles $\theta_1$, $\theta_2$, and $\theta_3$, specifying the angle between the preceding and posterior links. Since its motion is restricted to a single plane and it has three revolute joints, this type of robots is known as 3R planar robots.

As stated before, forward kinematics involves calculating the end-effector position, denoted as $p$, and orientation, represented by $\varphi$ (the angle relative to the world $x$-axis), based on the configuration of the robot, i.e., its joint angles. This is done using proper rigid body transformations (translations and rotations) that relate the world reference system to the end-effector coordinate system. Mathematically, the end-effector position and orientation of the robot is defined by the continuous functions:
\begin{equation}
	p=f(\theta_1,\theta_2,\theta_3) \quad \text{and} \quad \varphi=g(\theta_1,\theta_2,\theta_3)
\end{equation}
The computation of $p$ and $\varphi$ is done using PGA, because as stated in \ref{PGA}, PGA is well-suited to encode translations and rotations. Furthermore, as the forward kinematics problem does not involve the computation of spheres and circles, there is no necessity for the additional dimension and properties provided by CGA. The following code presents all the computation steps of the solution of the forward kinematics problem for an arbitrary 3R planar robot (where link lengths are also given as symbolic variables):

\begin{colorverbatim}
	GA([2,0,1])                             % PGA 2D algebra initialization
	syms angle_1 angle_2 angle_3 real      % joint angles
	syms length_1 length_2 length_3 real   % link lengths
	
	Point2PGA=@(x,y)dual(x*e1 + y*e2 + e3) % function def.: point embedding for PGA2D
	PGA2Point=@(P) [P(e23),-P(e13)]		      % function def.: point unembedding for PGA2D
	P0=Point2PGA(0,0)                      % point at the origin
	
	D1=exp(length_1/2*e13);                % translation rotor of the first link
	D2=exp(length_2/2*e13);                % translation rotor of the second link
	D3=exp(length_3/2*e13);                % Translation rotor of the third link
	
	R1=exp(-angle_1/2*e12);                % rotation rotor of the first angle
	R2=exp(-angle_2/2*e12);                % rotation rotor of the second angle
	R3=exp(-angle_3/2*e12);                % rotation rotor of the third angle
	
	R=R1*D1*R2*D2*R3*D3; 		                % end-effector rotor 
	P_ee=R*P0*~R;  			                       % end-effector position
	P_ee_kin=simplify(PGA2Point(P_ee));      % end-effector symbolic explicit kinematics
\end{colorverbatim}
\noindent which outputs the following solution:
\begin{colorverbatim}
	P_ee_kin =
	[length_2*cos(angle_1 + angle_2) + length_1*cos(angle_1) + length_3*cos(angle_1 + 
	angle_2 + angle_3), length_2*sin(angle_1 + angle_2) + length_1*sin(angle_1) + 
	length_3*sin(angle_1 + angle_2 + angle_3)]
\end{colorverbatim}
\noindent which coincides with the well-known expression for the forward kinematics of the 3R planar robot \cite{Siciliano08a}:
\begin{equation}\label{eq:traj}
	p=\left(\begin{array}{c}x \\ y \end{array}\right)=\left(\begin{array}{c} \ell_1\cos(\theta_1) +\ell_2\cos(\theta_1+\theta_2)+\ell_3\cos(\theta_1+\theta_2+\theta_3) \\ \ell_1\sin(\theta_1) +\ell_2\sin(\theta_1+\theta_2)+\ell_3\sin(\theta_1+\theta_2+\theta_3) \end{array}\right) \quad \text{and} \quad \varphi=\theta_1+\theta_2+\theta_3
\end{equation}
Therefore, the position and orientation of the robot's end-effector can be represented by the rotor encoding its forward kinematics, i.e., the rotor encoding the geometric transformation that relates the world reference system to the end-effector coordinate system.

\begin{figure}[h]
	\centering
	\includegraphics[width=2.5cm]{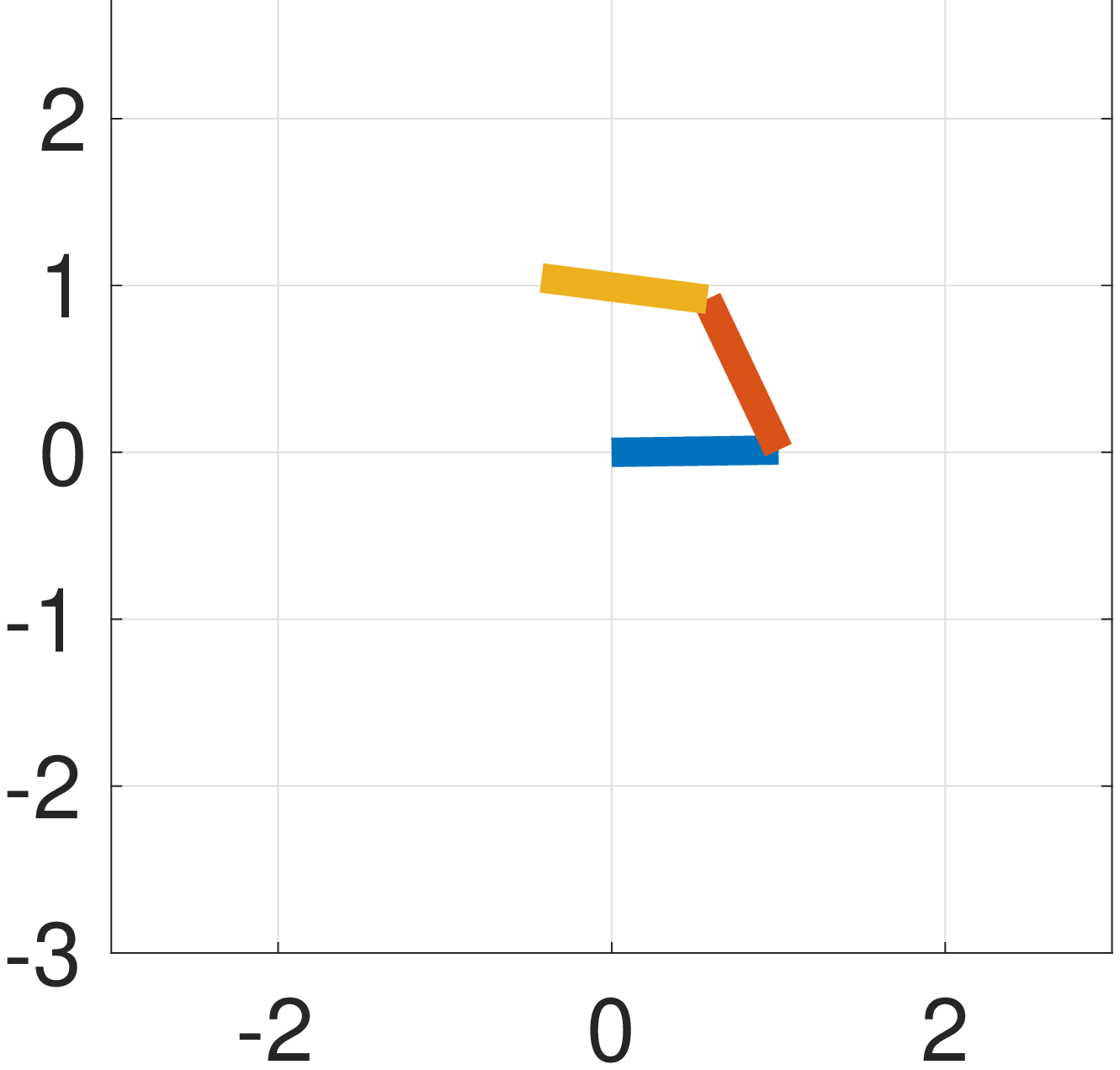}\includegraphics[width=2.5cm]{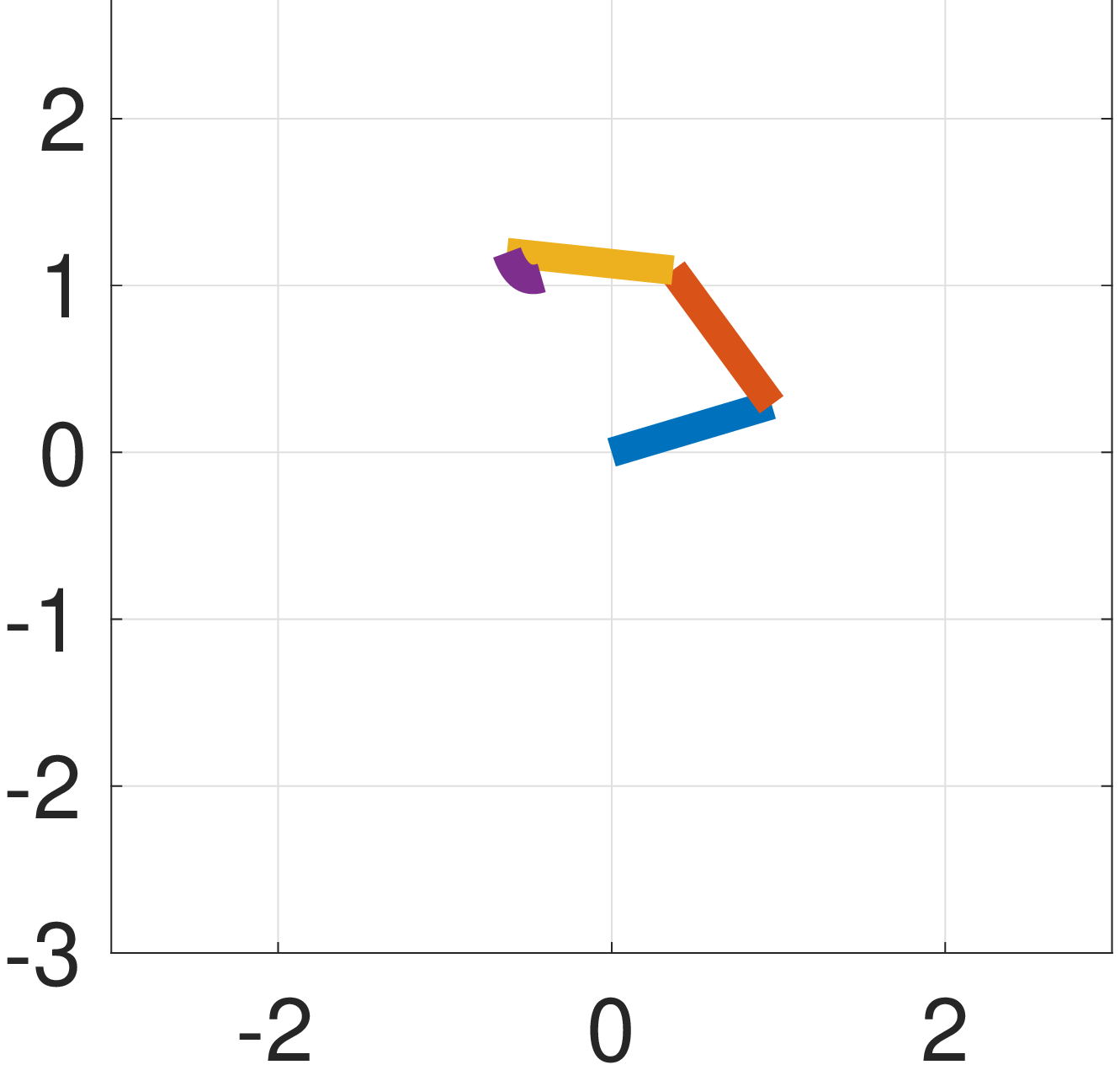}
	\includegraphics[width=2.5cm]{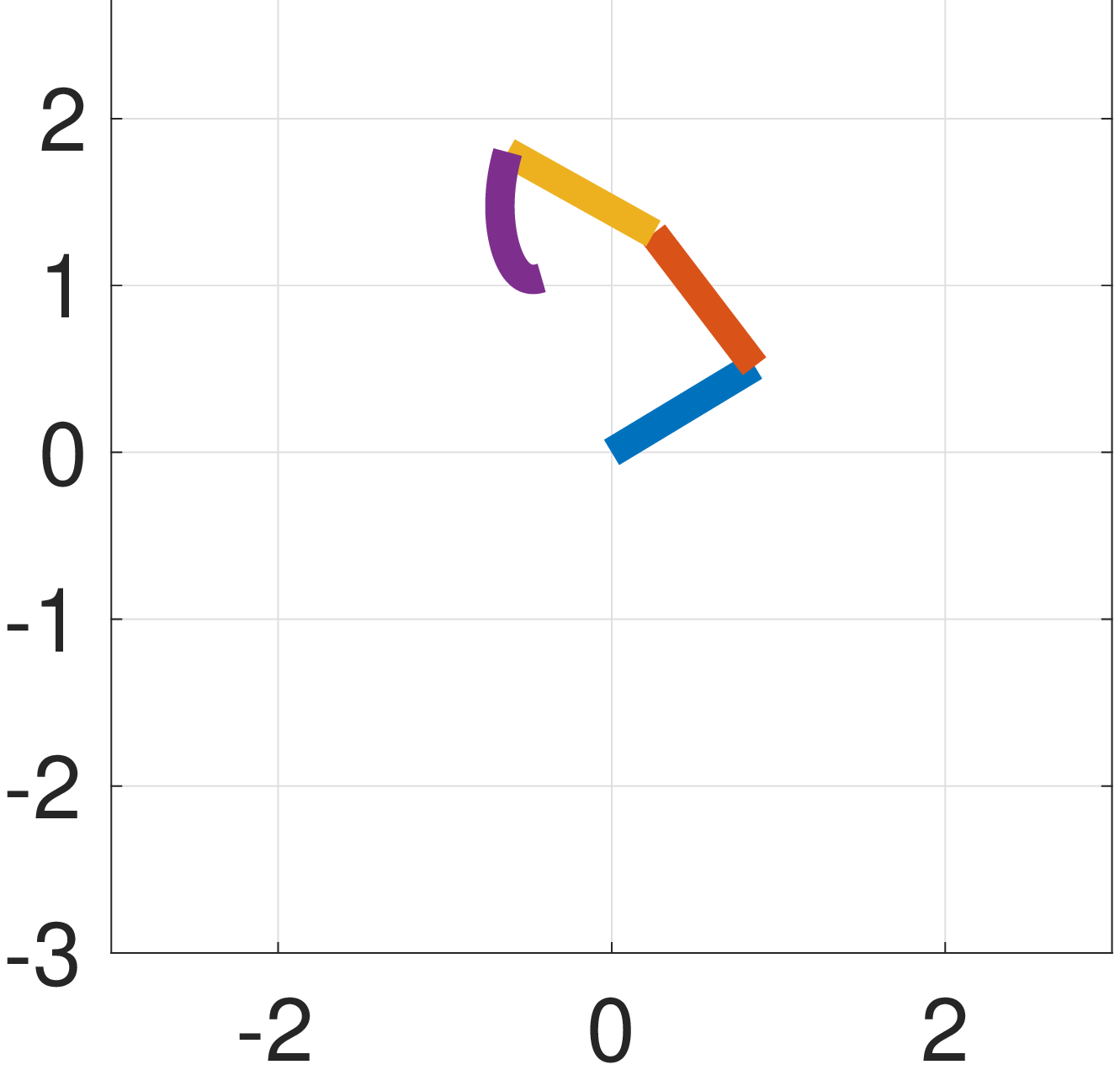}\includegraphics[width=2.5cm]{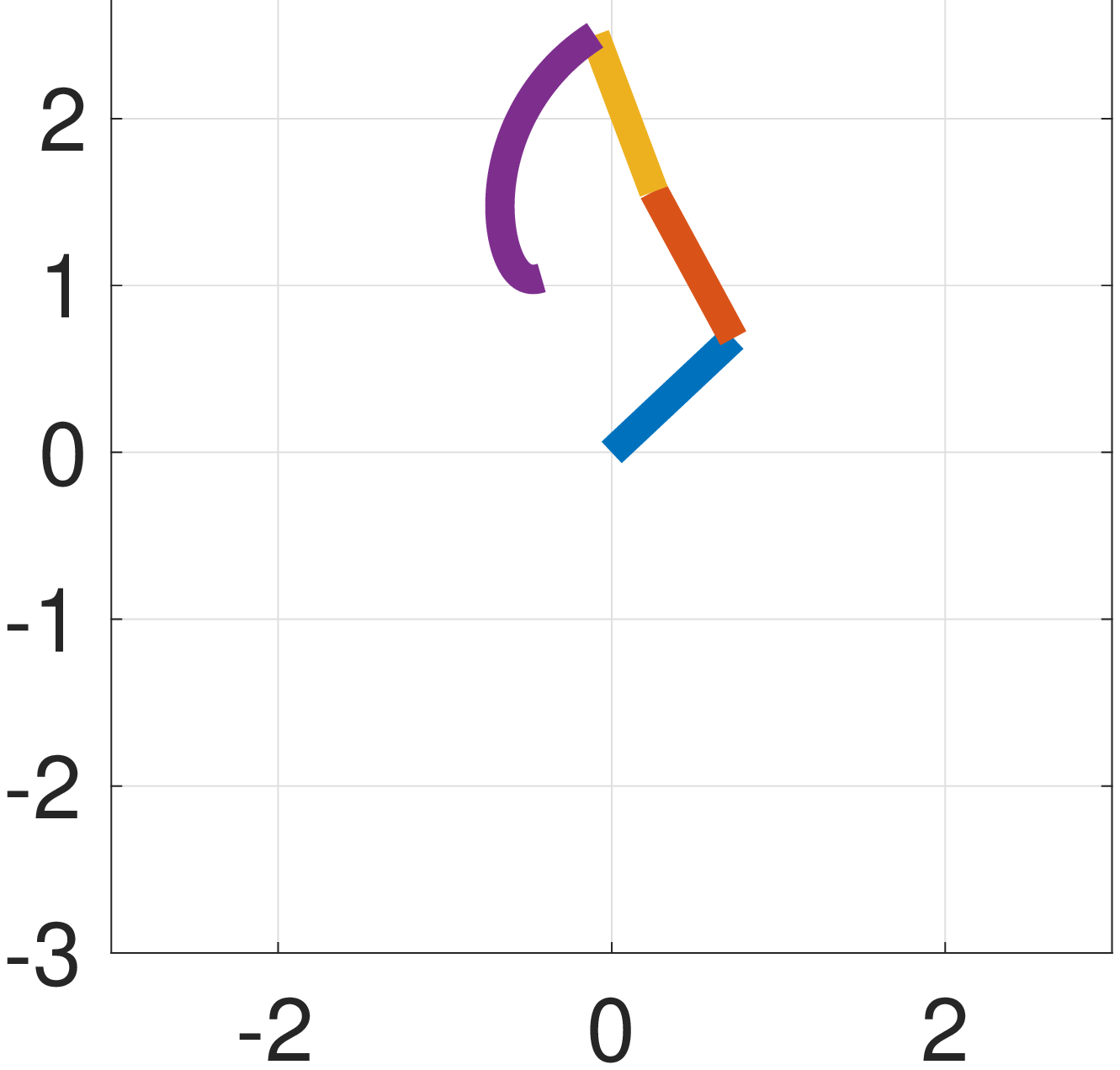}
	\includegraphics[width=2.5cm]{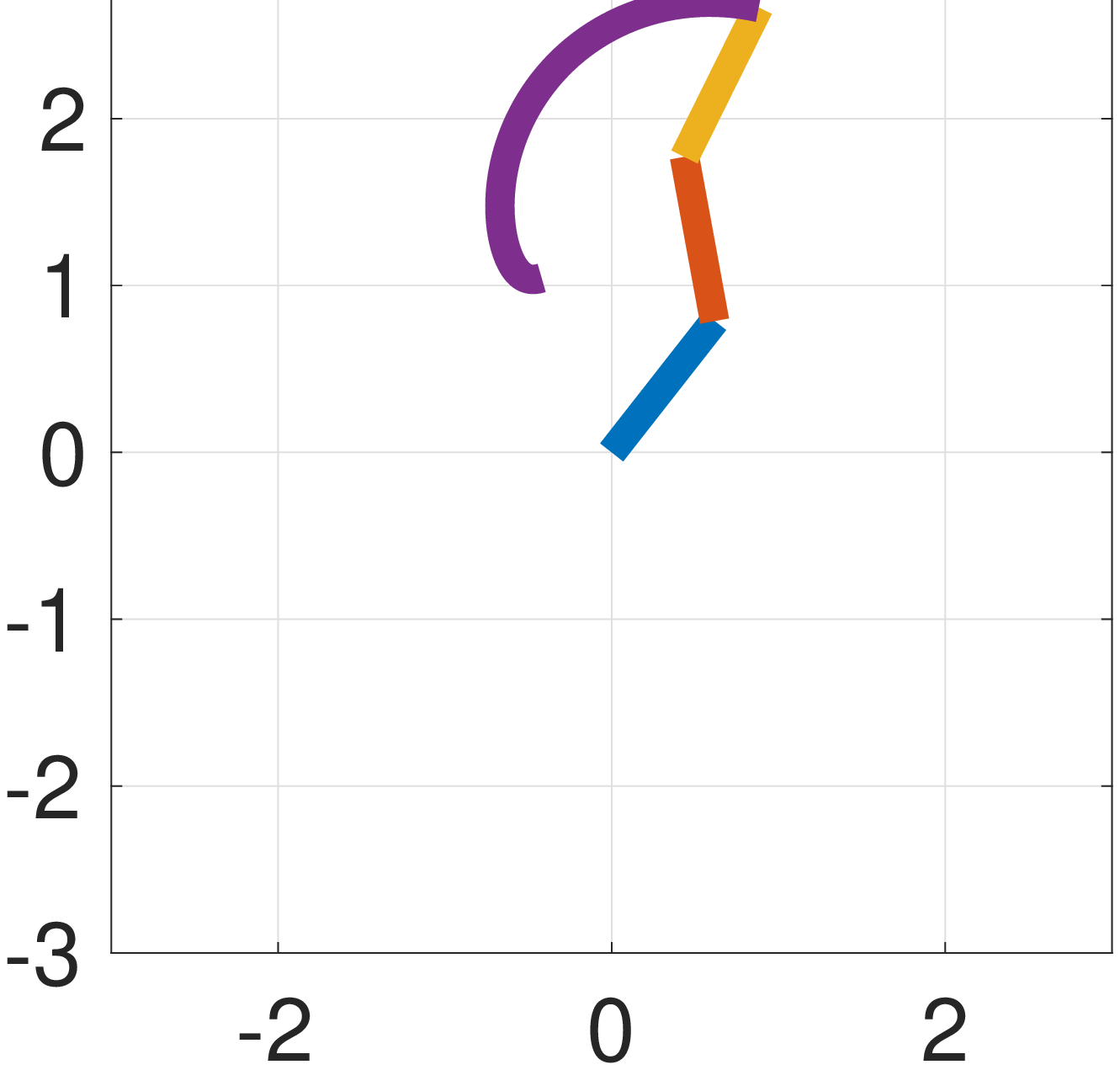}\includegraphics[width=2.5cm]{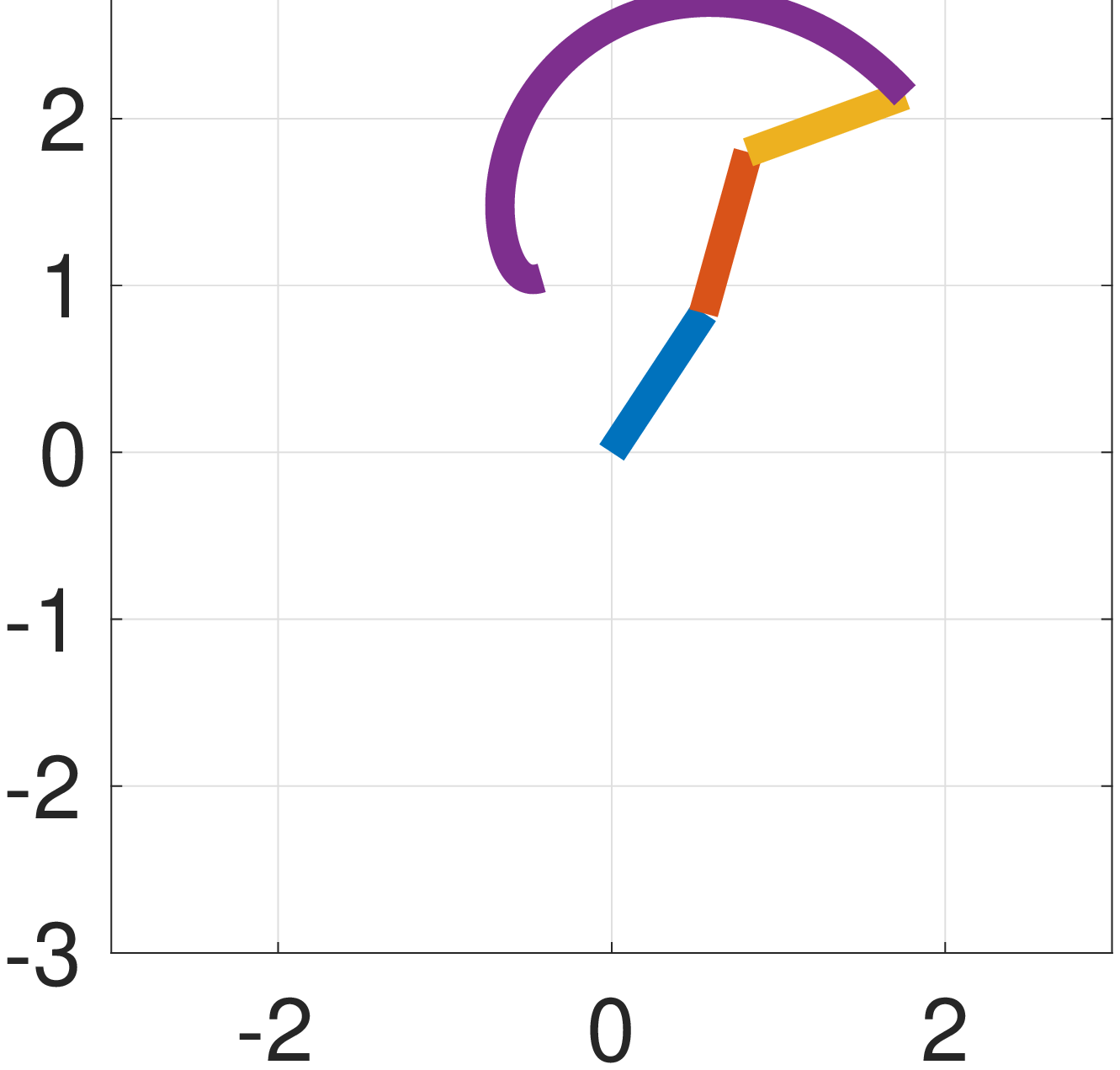}
	\includegraphics[width=2.5cm]{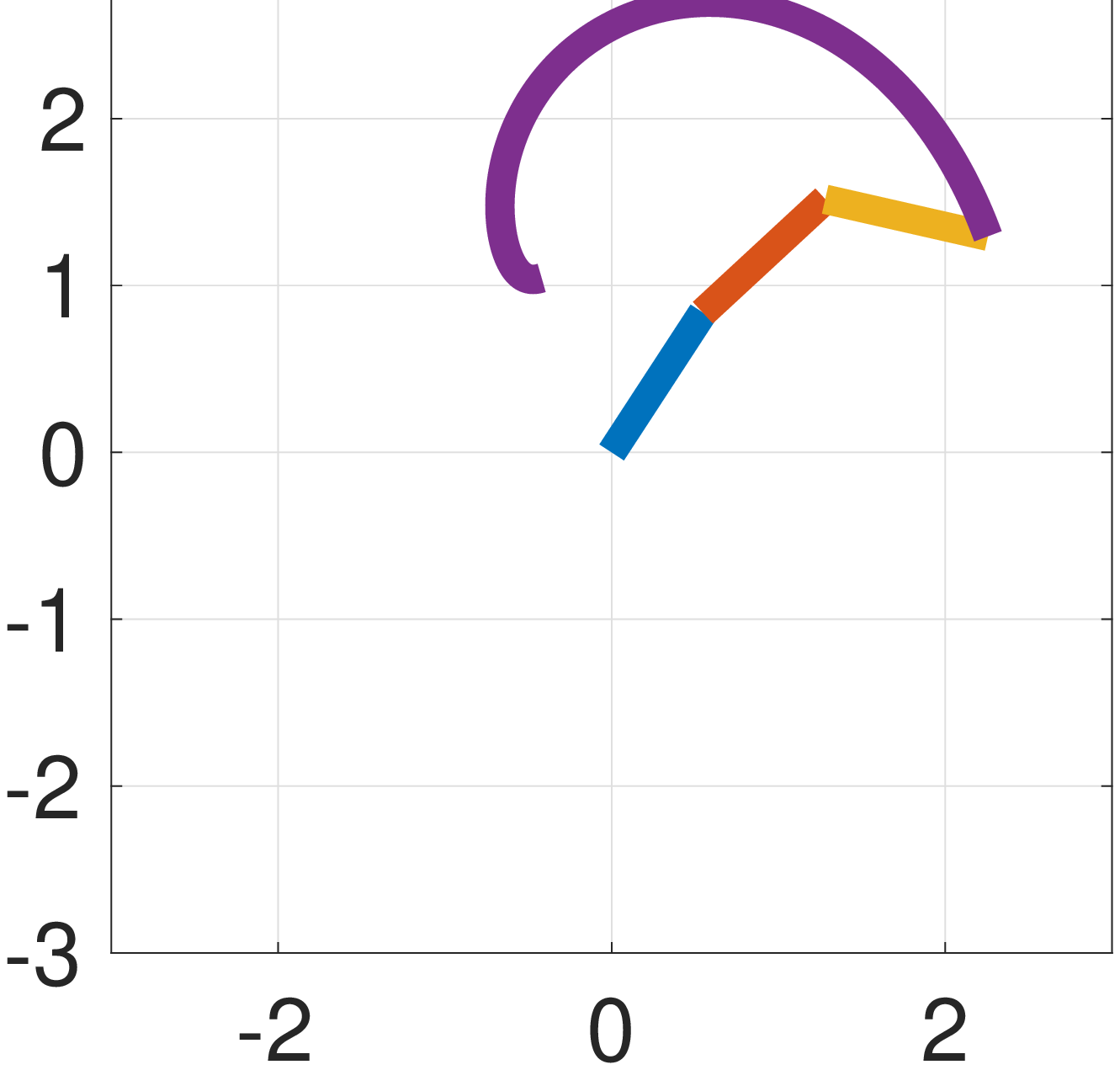}\\
	\includegraphics[width=2.5cm]{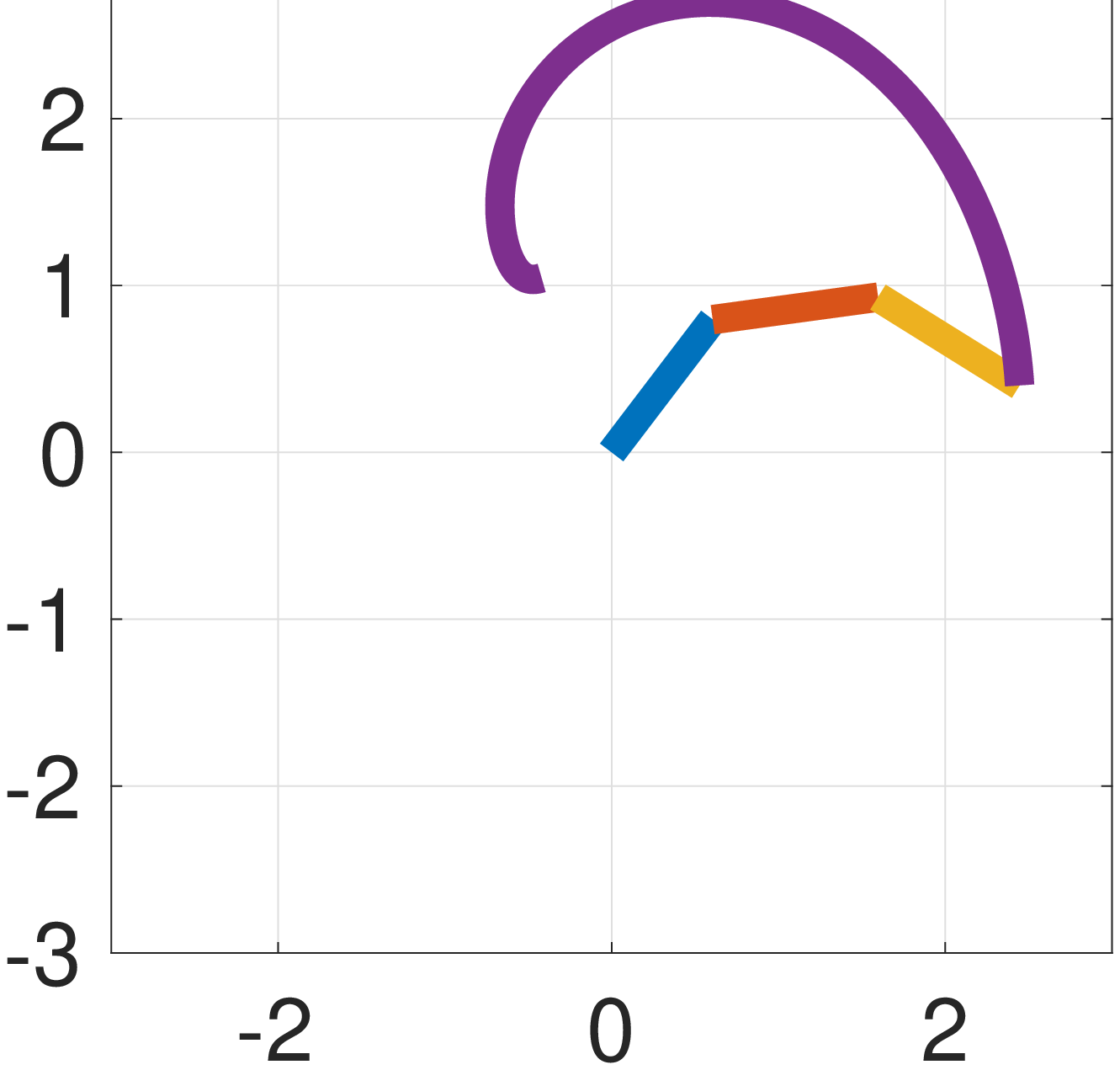}\includegraphics[width=2.5cm]{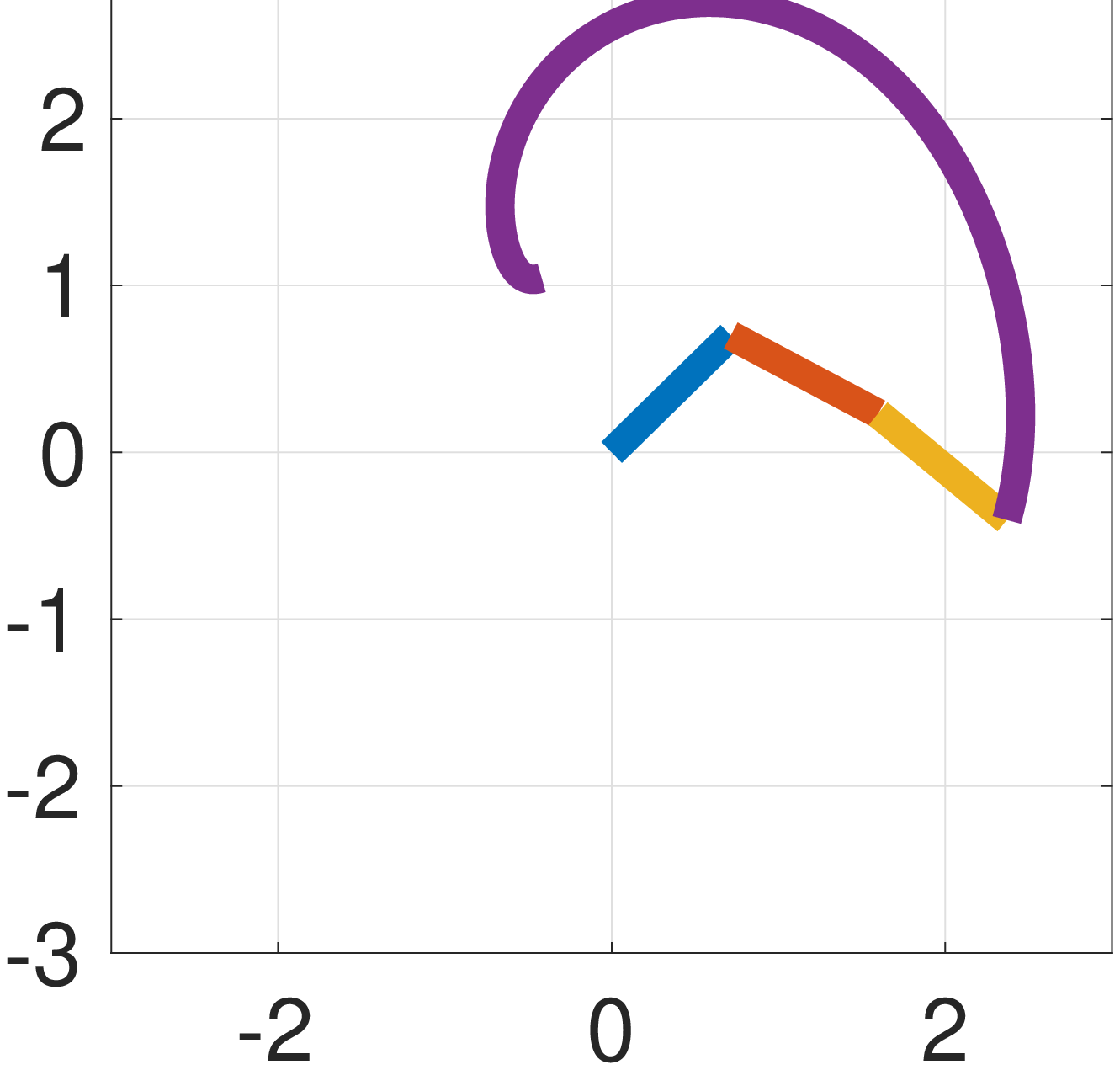}
	\includegraphics[width=2.5cm]{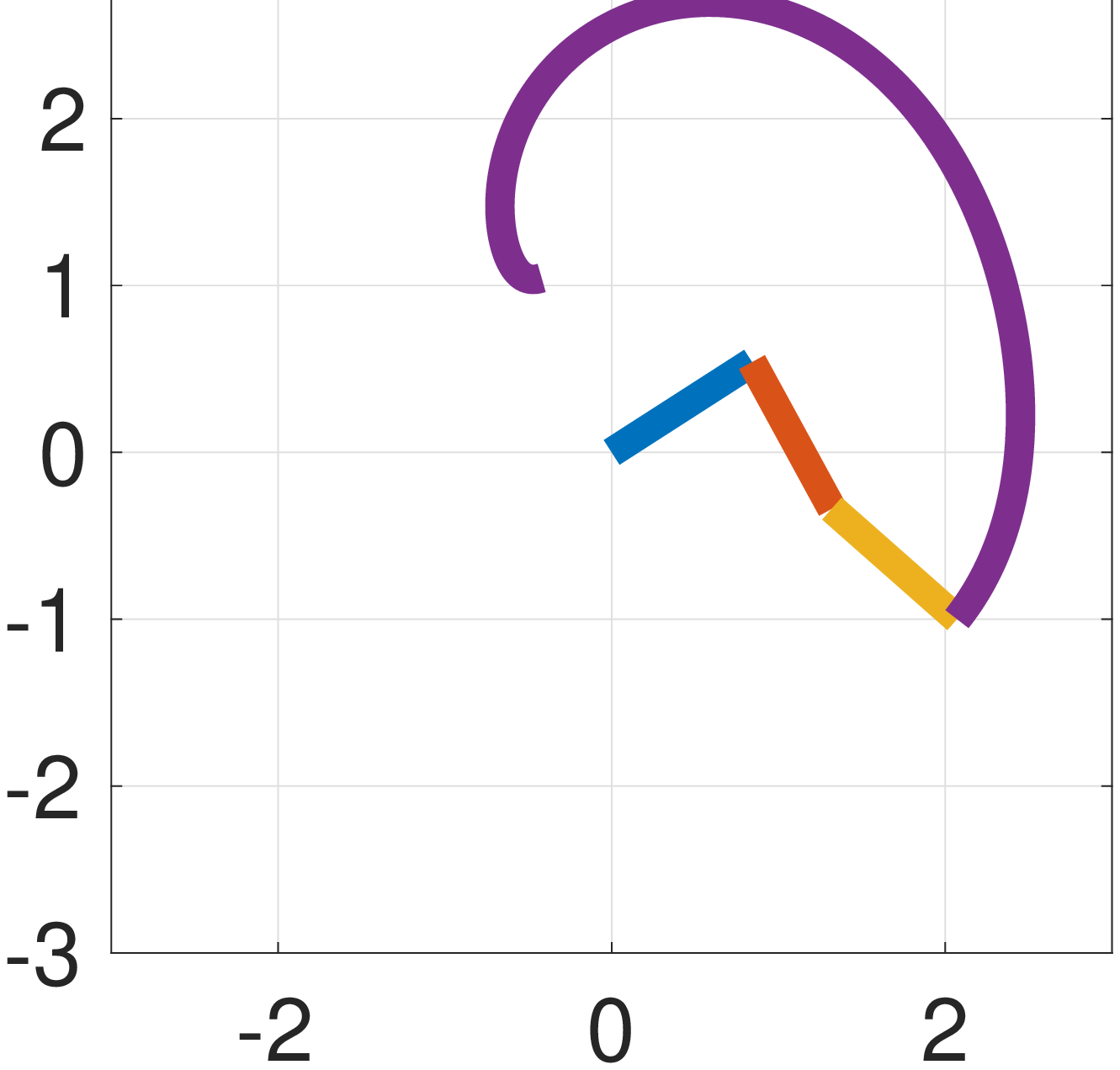}\includegraphics[width=2.5cm]{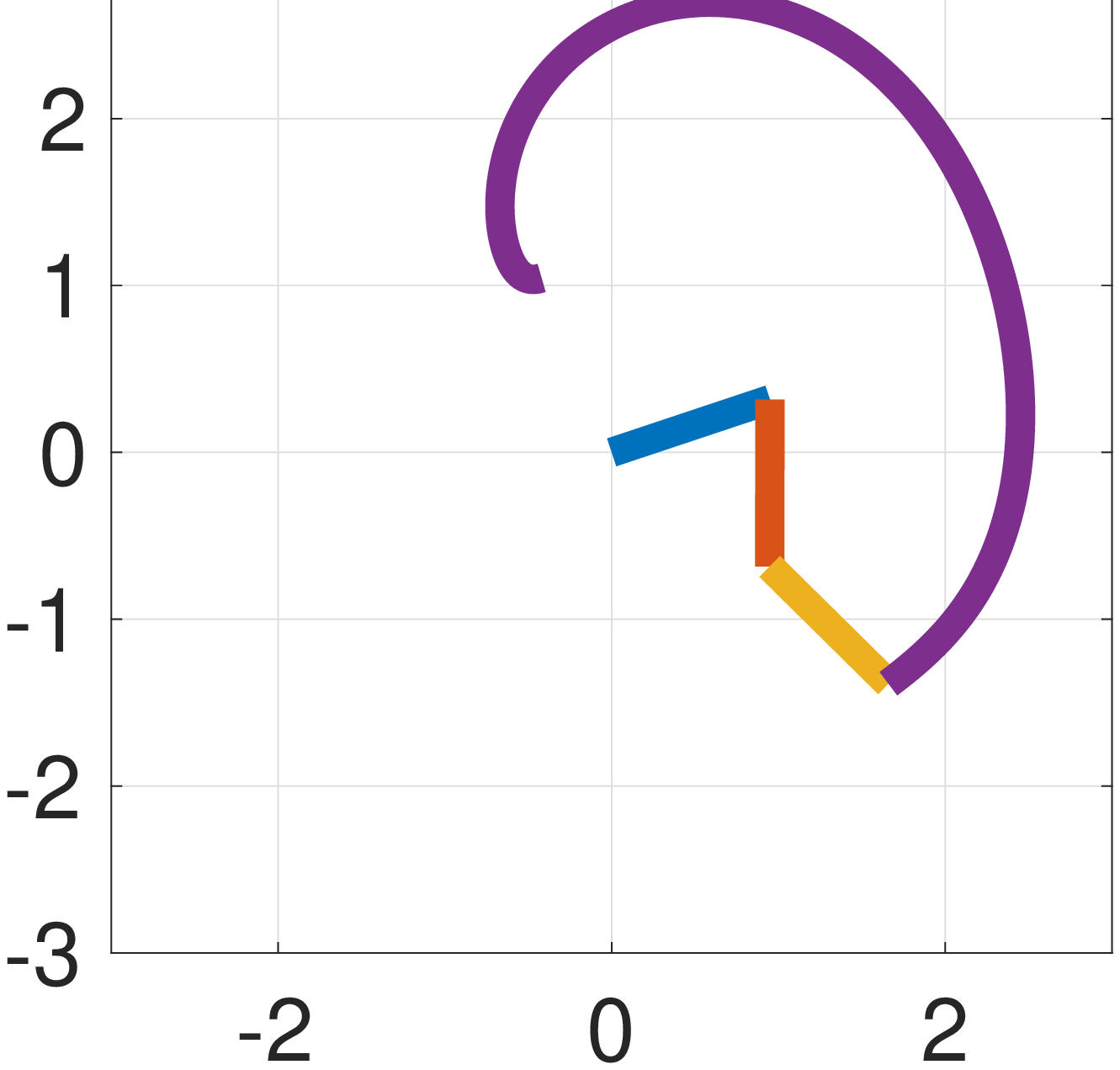}
	\includegraphics[width=2.5cm]{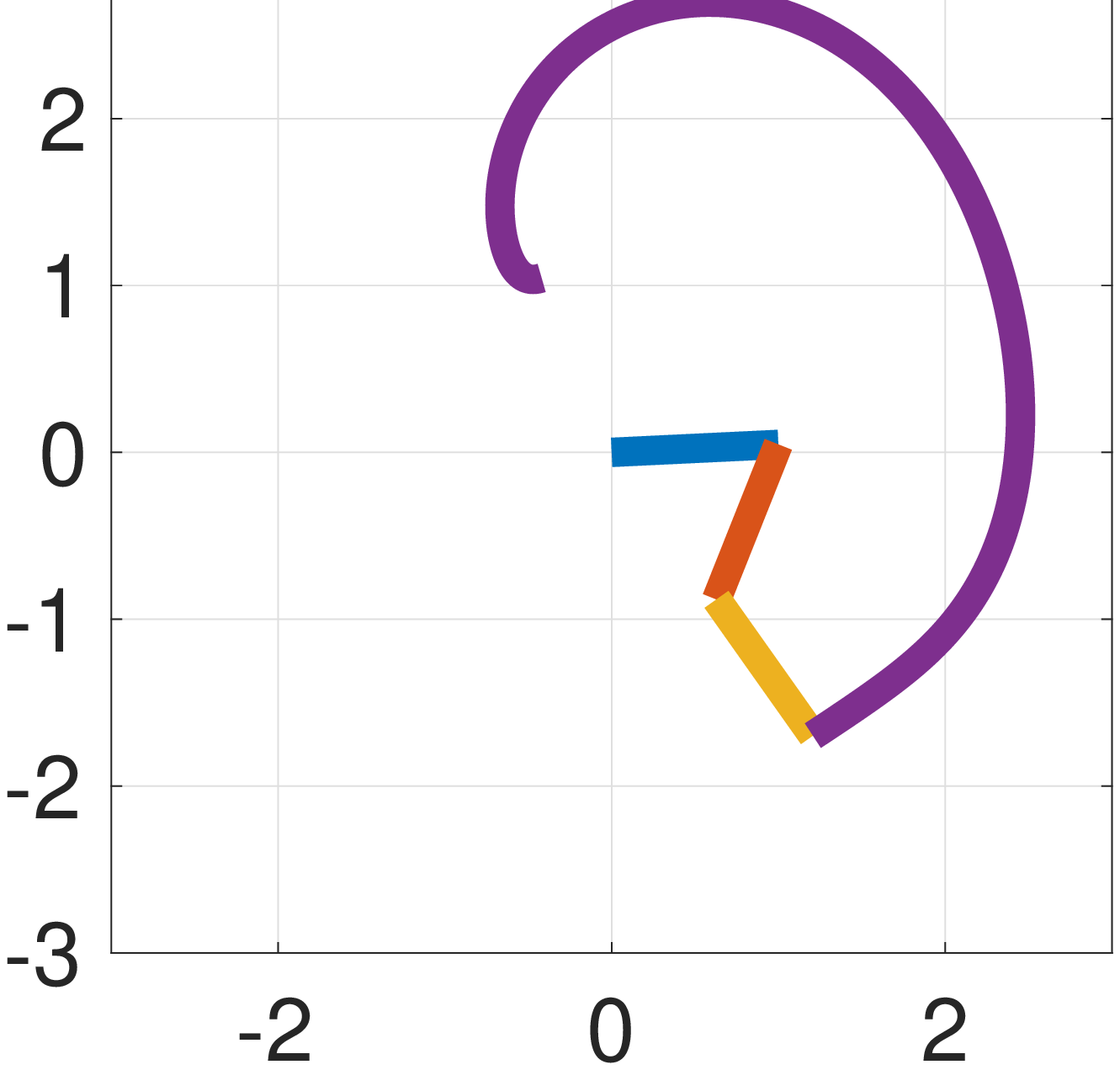}\includegraphics[width=2.5cm]{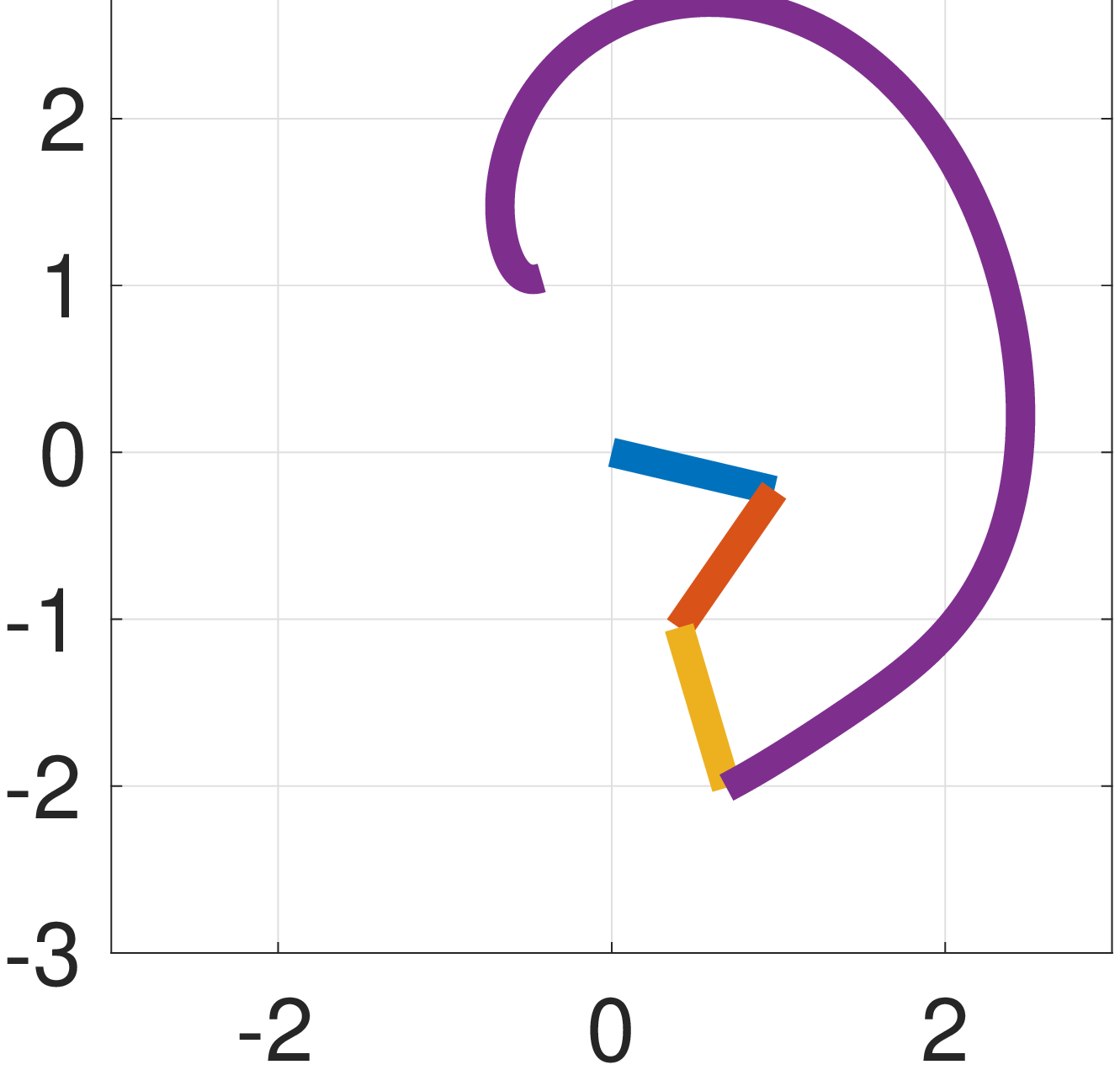}
	\includegraphics[width=2.5cm]{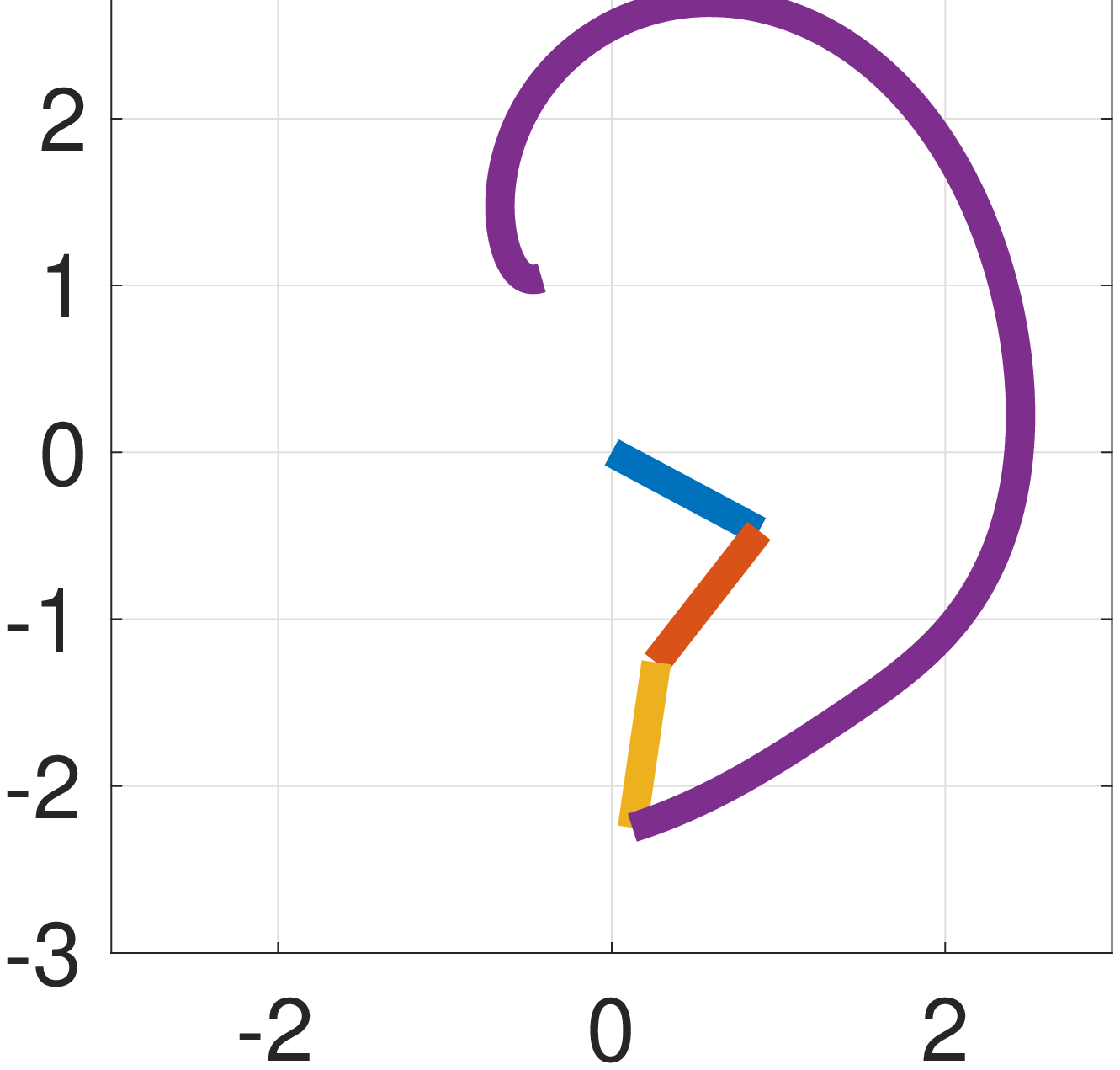}
	\caption{Trajectory for a 3R planar robot.\label{fig:2Drobotseq}}
\end{figure}

Finally, for every set of particular instances for the angles and lengths, the end-effector position and orientation can be determined via equation \eqref{eq:traj}. To illustrate the capabilities of $\mathtt{SUGAR}$, a trajectory has been generated varying all the angles within a given range. Fig. \ref{fig:2Drobotseq} shows a sequence of plots where the end-effector of the $3$R planar robot arm (first joint in blue, second joint in red, and third joint in yellow) performs a trajectory (in purple), computed with equation \eqref{eq:traj} given these varying values for the angles. 

\subsection{Inverse kinematics of a 6R robot}\label{ex:IK}

\begin{figure}[h!]
	\centering
	\includegraphics[width=7.5cm]{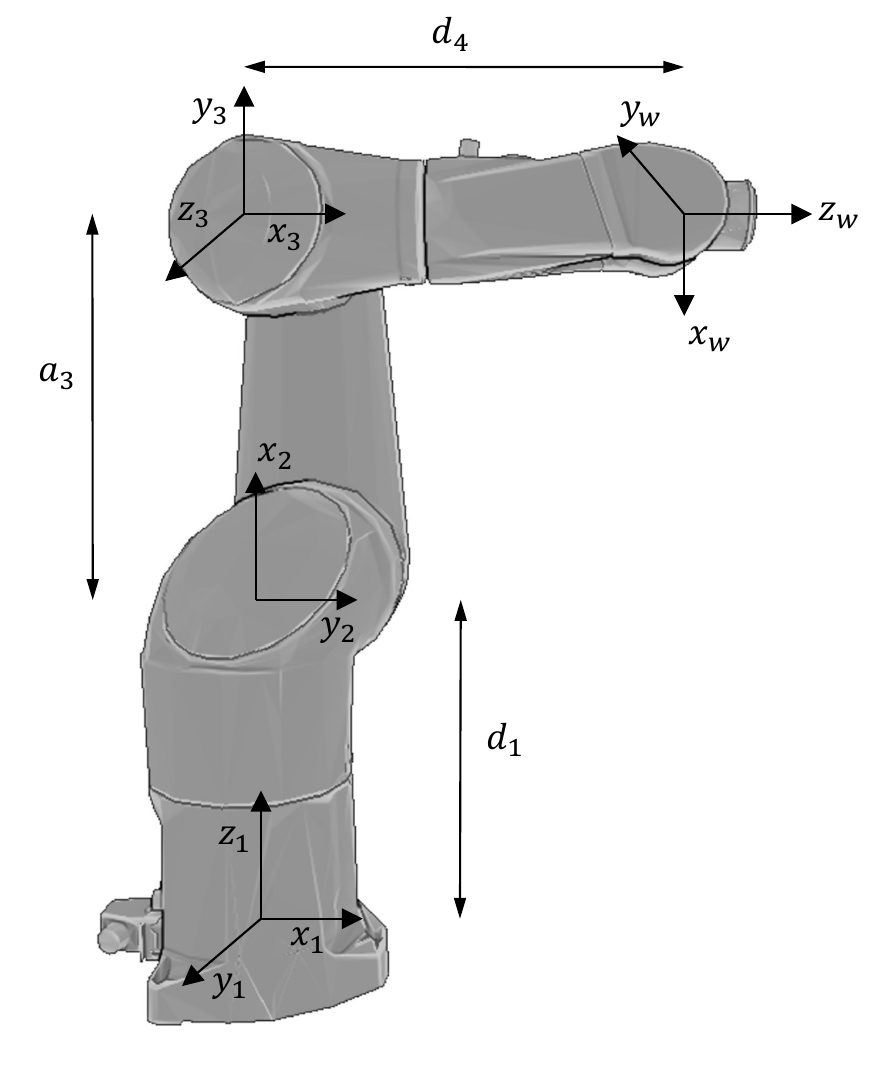}
	\caption{Schematic representation of a 6R robot.\label{fig:stauling}}
\end{figure}

Consider a robot arm as the one depicted in Fig.~\ref{fig:stauling}. This is an example of one of the most typical industrial serial robots. The objective of this example is to illustrate the user-friendly capabilities of $\mathtt{SUGAR}$, as well as the manipulation of CGA-based solution strategies using $\mathtt{SUGAR}$. For that, the kinematic model and solution equations are taken directly from the source \cite{Lav18} and implemented in $\mathtt{SUGAR}$. 

In particular, the robot consists of three rigid links with lengths $d_1$, $a_3$, and $d_4$ (named according to the Denavit-Hartemberg convention \cite{Siciliano08a}), and six revolute joints connected in series. As in the case of 3R planar robots, this type of robot is known as a 6R robot. The last three joints act on the same point, contributing only to the orientation of the robot's end-effector and not to its position. Similarly, the first three joints contribute to both the orientation and position of the robot's end-effector. Therefore, to determine the spatial position of the end-effector of the robot, only the three first joints are needed. Its joint variables are three angles denoted as $\theta_1$, $\theta_2$, and $\theta_3$.

As stated before, inverse kinematics consists of finding the set all joint angles for a given position $p$ and orientation $\varphi$ of the robot's end-effector. For this application example, and thanks to decoupling between position and orientation that the particular geometry of the robot allows, only the inverse position problem is considered. This means computing set of all values for $\theta_1,\theta_2$ and $\theta_3$ or which the end-effector, in that configuration, is at the specified position $p$.

Following the solution developed by Lavor et al. \cite{Lav18}, if $p_{0}$ denotes the point placed at the origin of the world frame, then two intermediate points $p_{1}$ and $p_{2}$ are needed to be found. The steps are:
\begin{itemize}
	\item The point $p_0$ is represented as a null vector of $\mathcal{G}_{4,1}$:
	\begin{equation}
		\bm{p}_0 = H(p_0) = \dfrac{1}{2}p_0^2n_\infty+n_0 +p_0.
	\end{equation}
	\item The null vector $\bm{p}_0$ is translated in the direction $z_1$ of the world frame an amount equal to the first link length, i.e., $d_1$:
	\begin{equation}
		\bm{p}_{1} = T_{z_1}\bm{p}_0\widetilde{T}_{z_1},
	\end{equation}
	where $T_{z_1} = 1 - \frac{z_1\wedge n_\infty}{2}$.
	\item The inner representation of a sphere centered at $\bm{p}_1$ and with radius $a_3$ is computed:
	\begin{equation}\label{S1}
		\bm{s}_{1}^\ast = \bm{p}_{1} - \dfrac{1}{2}a_{3}^{2}e_{\infty}
	\end{equation}
	\item The inner representation of a sphere centered at $\bm{p}$ (the null vector representation of the given position $p$) and with radius $d_4$ is computed:
	\begin{equation}\label{S2}
		\bm{s}_{2}^\ast = \bm{p} - \dfrac{1}{2}d_{4}^{2}e_{\infty}
	\end{equation}
	\item The output representation of a plane passing through $\bm{p}_{0}$, $\bm{p}_{1}$ and $\bm{p}$ is computed:
	\begin{equation}\label{plane1}
		\bm{\pi} = \bm{p}_{0}\wedge \bm{p}_{1}\wedge \bm{p}_{w}\wedge n_{\infty}
	\end{equation}
	\item The intersection of the plane and two spheres is computed:
	\begin{equation}\label{B2}
		B = (\bm{s}_{1}^{\ast}\wedge \bm{s}_{2}^{\ast}\wedge\bm{\pi}^\ast)^\ast,
	\end{equation}
	which is a bivector and, therefore, it represents a pair of points. In particular, $B = \bm{b}_{1}\wedge \bm{b}_{2}$ for some null vectors $\bm{b}_{1}$ and $\bm{b}_{2}$. 
	\item Null vectors $\bm{b}_{1}$ and $\bm{b}_{2}$ are extracted from $B$: 
	\begin{equation}\label{ExtractionP2}
		\begin{split}
			\bm{b}_{1} &= -\widetilde{P}(B\cdot n_\infty)P\\
			\bm{b}_{2} &= P(B\cdot n_\infty)\widetilde{P}
		\end{split}
	\end{equation}
	where $P$ denotes the projector operator defined as:
	\begin{equation}\label{Projector}
		P = \dfrac{1}{2}\biggl(1+\dfrac{B}{\sqrt{B\widetilde{B}}}\biggr)
	\end{equation}
	\item Null vector $\bm{p}_{2}$ is equal to any of the recovered null vectors $\bm{b}_{i}$, $i=1,2$, i.e., there are two different solutions.
	\item For each value of $\bm{p}_{2}$, three auxiliary lines are computed:
	\begin{align}
		\ell_{1} &= \bm{p}_{0}\wedge \bm{p}_{1}\wedge n_{\infty}\label{l1}\\
		\ell_{2} &= \bm{p}_{1}\wedge \bm{p}_{2}\wedge n_{\infty}\label{l2}\\
		\ell_{3} &= \bm{p}_{2}\wedge \bm{p}\wedge n_{\infty}\label{l3}
	\end{align}
	\item The joint angles are calculated:
	\begin{align}
		\theta_{1} &=\angle(x_{1},\pi)\\
		\theta_{2} &=\angle(\ell_{1},\ell_{2})\\
		\theta_{3} &=\angle(\ell_{2},\ell_{3})
	\end{align}
	where $x_1$ is the $x$-axis of the world reference system.
\end{itemize}

\begin{figure}[h]
	\centering
	\includegraphics[scale=0.65]{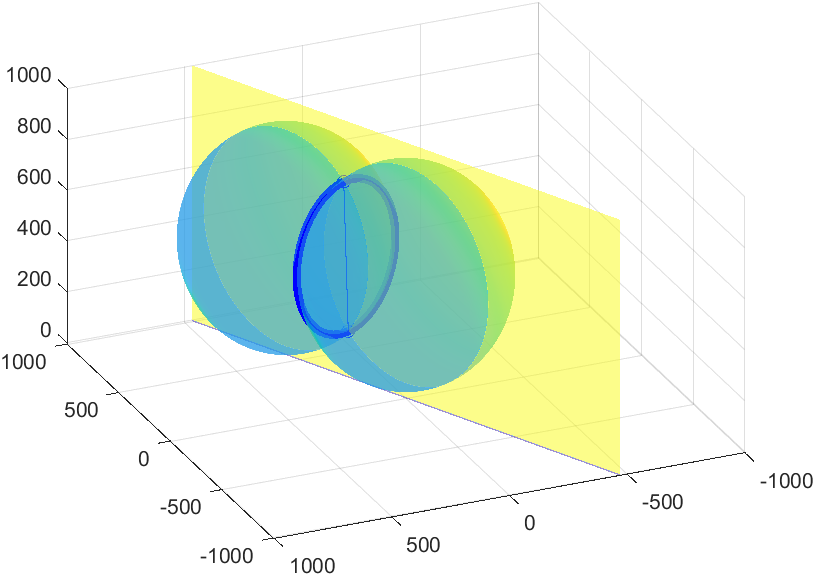}
	\caption{Geometric entities generated in the resolution of the inverse kinematics problem for the Stäubli robot and their intersections as generated in $\mathtt{SUGAR}$\label{plot_code}}
\end{figure}

The following code implements this solution in $\mathtt{SUGAR}$:

\begin{colorverbatim}
	CGA(3)                                    % define 3D conformal geometric algebra
	d1 = 480;                                % define the Staübli robot link lengths
	a3= 425;
	d4 = 425;
	
	q_d = [0.4375,0.8590,1.5040];            % desired configuration
	pos_d = [561.8479,262.7685,455.0104];    % associated desired position of the end-
	% effector
	M_pos_d = pos_d(1)*e1+pos_d(2)*e2+pos_d(3)*e3 % multivector of the desired position
	POS_D = push(M_pos_d);                        % push of the desired position (eq. 27)
	
	P0 = n0;                                      % compute p_0 (point at the origin)
	T1 = make_translation(d1,e3);                 % define the rotor for the translation
	P1 = T1.reverse*P0*T1;                        % compute p_1 (translation of p_1) (eq. 28)
	
	plane = P0.^P1.^POS_D.^ni;                    % plane passing by p_0, p_1 and pos_d (eq. 31)
	
	a = 0.5*(a3^2);                               % parameters from the link lengths
	d = 0.5*(d4^2);
	
	S1 = P1 - a*ni;                               % sphere center = p_1 and radius = a3 (eq. 29)
	S2 = POS_D - d*ni;                            % sphere center = pos_d and radius = d4 (eq. 30)
	
	C = S1.^S2;                                   % intersection of the spheres S1 and (eq. 32)
	% S2, results in a circle
	B = dual(plane).^clean(C);                    % intersection of the plane plane and (eq. 32)
	% circle C, results in a pair of points
	S1.plot()                                     % plot of all the geometric entitites
	hold on                                       % and their intersections
	
	S2.plot()
	plane.normalize().plot()
	C.plot()
	B.plot();
	xlim([-1000,1000])
	ylim([-1000,1000])
	hold off
	
	P2_sol1 = B.info.P1;                          % extract first point P2 from B (eq. 33)
	P2_sol2 = B.info.P2;                          % extract second point P2 from B (eq. 33)
	% continue with only one P2, the other
	% is completely analagous
	normal_plane = dual(plane).info.n;            % normal vector to plane plane
	normal = normal_plane(e1+e2+e3);              % normal vector as GA vector
	x = [1,0,0];
	
	% first joint variable
	
	q1 = acos(dot(normal,x)/(norm(x)*norm(normal)))-pi/2; % (38)
	
	l1 = P0.^P1.^ni;                              % auxiliary line l1 (eq. 35)
	l2_sol1 = P1.^P2_sol1.^ni;                    % auxiliary line l2 (eq. 36)
	l3_sol1 = P2_sol1.^POS_D.^ni;                 % auxiliary line l3 (eq. 37)
	
	L11 = l1*l1;                                  % module of line l1
	L11_scalar = L11(G0);                         % module as double
	L22_sol1 = l2_sol1*l2_sol1;                   % module of line l2
	L22_scalar_sol1 = L22_sol1(G0);               % module as double
	L33_sol1 = l3_sol1*l3_sol1;                   % module of line l3
	L33_scalar_sol1 = L33_sol1(G0);               % module as double
	
	L12_sol1 = l2_sol1.*l1;                       % angle lines l1 and l2
	L12_scalar_sol1 = L12_sol1(G0);               % angle as double 
	
	L23_sol1 = l2_sol1.*l3_sol1;                  % angle lines l2 and l3
	L23_scalar_sol1 = L23_sol1(G0);               % angle as double
	
	% second and third joint variables
	
	q2_sol1 = acos((L12_scalar_sol1)/(sqrt(L11_scalar)*sqrt(L22_scalar_sol1))); % (eq. 39)
	q3_sol1 = acos((L23_scalar_sol1)/(sqrt(L22_scalar_sol1)*sqrt(L33_scalar_sol1))); %(eq. 40)
	
	q_proposed_sol1 = [q1,q2_sol1,q3_sol1]        % solution configuration
\end{colorverbatim}
\noindent which outputs the following solution:
\begin{colorverbatim}
	q_proposed_sol1 =
	0.4375
	0.8590
	1.5040
\end{colorverbatim}
\noindent that matches the input desired robot configuration. In addition, as demonstrated in the provided code, the translation of the theoretical solution presented by Lavor et al. \cite{Lav18} can be easily implemented in $\mathtt{SUGAR}$ (in fact, it is completely straightforward). This, in turns, shows that the user-friendly part of SUGAR is justified. Furthermore, aside from the straightforward implementation of any computation step in GA, PGA and CGA, the CGA module incorporates a visualization tool that is essential to understand the complex geometric manipulations typically associated with mathematical and engineering applications using CGA. Fig. \ref{plot_code} depicts, as an illustrative example, the plot generated during the execution of the provided code.

\subsection{Power systems analysis}\label{ex:PSA}

This example illustrates the capability of $\mathtt{SUGAR}$ for operating with matrices of multivectors, and in particular, for calculating the inverse of a matrix of multivectors. Such operation, which in general is rarely found in other existing computational libraries (or even inexistent for symbolic computing), can be exploited for example when working with  electrical systems modelled using a GA approach \cite{Vel23}.

Unbalanced three-phase electrical systems can be represented using GA, where voltages and currents are multivectors and are related by geometric impedances with the form
\begin{equation}
	z(s)=z_\mathrm{av}(s)e_0+z_\mathrm{unI}(s)e_1+z_\mathrm{unR}(s)e_2
\end{equation}
where $z_\mathrm{av}(s),z_\mathrm{unI}(s)$ and $z_\mathrm{unR}(s)$ are transfer functions. In case of balanced impedances $z_\mathrm{unI}(s)=z_\mathrm{unR}(s)=0$. See more details in \cite{DVZM2024}.

\begin{figure}[h]
	\begin{center}
		\scalebox{1}{\begin{circuitikz} 
\draw (0,0) node [oscillator,scale=1.0](slack) {} 
	(slack.e) to [tmultiwire,-*]++(0.75,0) node [anchor=north] {$1$} coordinate (N1) 
	to [generic,l=$z_{12}(s)$,-*]++(1.5,0) 	node [anchor=south] {$2$} coordinate (aux)
	to [generic,l=$z_{23}(s)$,-*]++(1.5,0) 	node [anchor=west] {$3$}
	to [short]++(0,-0.5) to [generic,l=$z_{L3}(s)$]++(0,-0.65) 
	(aux) to [short]++(0,-0.5) to [generic,l=$z_{L2}(s)$]++(0,-0.65)
	(N1) to [short]++(0,1) to [generic,l=$z_{13}(s)$]++(3,0) to [short]++(0,-1)
	(-0.25,0.25) node [anchor=south] {$v$}
;

\end{circuitikz} }
	\end{center}
	\caption{A three-phase unbalanced electrical power network.}
	\label{fig_circuit}
\end{figure}
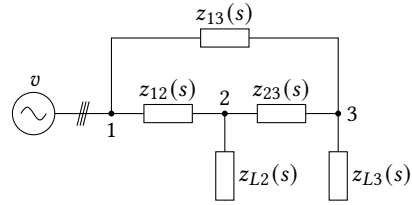

Figure \ref{fig_circuit} shows the equivalent circuit of a three-phase electrical power grid with an ideal power source (slack), three lines connecting nodes 1, 2, and 3, and two consumers at nodes 2 and 3. The geometric impedances $z_{12}(s),z_{13}(s),z_{23}(s)$ represent the dynamics and dissipation of the connection lines, $z_{L2}(s),z_{L3}(s)$ represent the electrical loads of consumers, and $v_s$ is the multivector representing the power source voltage.

The nodal analysis is one of the most used approaches for the analysis of power networks \cite{Dimo1975} and allows us to find the voltages at all the nodes of the circuit by solving a matrix equation. The method is based on creating the admittance matrix, $Y$, of the circuit, where the matrix elements are composed of the circuit admittances defined as
\begin{equation}
	y(s)=z^{-1}(s).
\end{equation}

For the example in Figure \ref{fig_circuit}, the matrix equation results in
\begin{equation} 
	\underbrace{\begin{pmatrix}
			y_{12}(s)+y_{13}(s)&-y_{12}(s)&-y_{13}(s)&1\\
			-y_{12}(s)&y_{12}(s)+y_{23}(s)+y_{L2}(s)&-y_{23}(s)&0\\
			-y_{13}(s)&-y_{23}(s)&y_{13}(s)+y_{13}(s)+y_{L3}(s)&0\\
			1&0&0&0
	\end{pmatrix}}_{Y}
	\begin{pmatrix}v_1\\v_2\\v_3\\i_s
	\end{pmatrix}=
	\begin{pmatrix}0\\0\\0\\v_s
	\end{pmatrix}\label{eq_MatrixEquation}
\end{equation}
where the system unknowns are the node voltages, $v_1,v_2,v_3$, and the current of the power source, $i_s$.

Consider the power network described in a per-unit (adimensional) system, with $v_s=v_\alpha(t)e_0+v_\beta(t)e_1+v_\alpha(t)e_2+v_\beta(t)e_{12}$ and the line\footnote{Notice that the lines has been considered balanced.} and load geometric impedances with values
\begin{align}
	z_{12}(s)&=(0.02s+0.01)e_0\\
	z_{13}(s)&=(0.04s+0.02)e_0\\
	z_{23}(s)&=(0.02s+0.01)e_0\\
	z_{L2}(s)&=0.5e_0-0.0289e_1+0.05e_2\\
	z_{L3}(s)&=0.4e_0-0.11559e_1-0.1e_2
\end{align}
with the aim of finding the voltage dynamics in node 2.

The solution of \eqref{eq_MatrixEquation} can be easily done in $\mathtt{SUGAR}$:
\begin{colorverbatim}
	syms s;                                           % symbolic Laplace variable
	syms va vb;                                      % symbolic voltage variables
	GA([2 0 0]);                                     % GA definition
	
	r12=0.01; l12=0.02; 	                            % line 12 parameters
	r13=0.02; l13=0.04; 	                            % line 13 parameters
	r23=0.01; l23=0.02; 	                            % line 23 parameters
	R2av=0.5; R2unI=-0.0289; R2unR=0.05;  	          % load 2 parameters
	R3av=0.4; R3unI=-0.1155; R3unR=-0.1;  	          % load 3 parameters
	vs=va*e0+vb*e1+va*e2+vb*e12;                     % slack voltage
	
	z12=(r12+l12*s)*e0;                              % line 12 impedance definition (eq. 48)
	z13=(r13+l13*s)*e0;                              % line 13 impedance definition (eq. 49)
	z23=(r23+l23*s)*e0;                              % line 21 impedance definition (eq. 50)
	zL2=R2av*e0+R2unI*e1+R2unR*e2;                   % load 2 impedance definition (eq. 51)
	zL3=R3av*e0+R3unI*e1+R3unR*e2;                   % load 3 impedance definition (eq. 52)
	
	y12=clean(inv(z12));                             % admittance calculation (eqs. 46 and 48)
	y13=clean(inv(z13));                             % admittance calculation (eq. 46 and 49)
	y23=clean(inv(z23));                             % admittance calculation (eq. 46 and 50)
	yL2=clean(inv(zL2));                             % admittance calculation (eq. 46 and 51)
	yL3=clean(inv(zL3));                             % admittance calculation (eq. 46 and 52)
	Y=[y12+y13 -y12 -y13 1*e0;                       % admittance matrix (eq. 47)
	-y12 y12+y23+yL2 -y23 0*e0;
	-y13 -y23 y13+y23+yL3 0*e0;
	1*e0 0*e0 0*e0 0*e0];
	
	x=inv(Y)*[0*e0;0*e0;0*e0;vs]                     % solving the system
	
	v2=x(2);                                         % MV voltage at node 2
	[Nv2e0,Dv2e0]=numden(v2(1))                      % polynomials of TF with e0
	[Nv2e1,Dv2e1]=numden(v2(2))                      % polynomials of TF with e1
	[Nv2e2,Dv2e2]=numden(v2(3))                      % polynomials of TF with e2
	[Nv2e12,Dv2e12]=numden(v2(4))                    % polynomials of TF with e12
\end{colorverbatim}
\noindent providing the polynomials of the transfer functions in the multivector $v_2$:
\begin{colorverbatim}
	Nv2e0=14595547*va-62633*vb+1596858*s*va-126698*s*vb+50148*s^2*va+440*s^3*va...
	-2876*s^2*vb-24*s^3*vb
	Dv2e0=16*s^4+2592*s^3+136080*s^2+2552384*s+15052095
	
	Nv2e1=14651277*vb-62553*va-126378*s*va+1706478*s*vb-2556*s^2*va-24*s^3*va...
	+46428*s^2*vb+360*s^3*vb
	Dv2e1=16*s^4+2592*s^3+136080*s^2+2552384*s+15052095 
	
	Nv2e2=14595547*va-62633*vb+1596858*s*va-126698*s*vb+50148*s^2*va+440*s^3*va...
	-2876*s^2*vb-24*s^3*vb
	Dv2e2=16*s^4+2592*s^3+136080*s^2+2552384*s+15052095
	
	Nv2e12=14651277*vb-62553*va-126378*s*va+1706478*s*vb-2556*s^2*va-24*s^3*va...
	+46428*s^2*vb+360*s^3*vb
	Dv2e12=16*s^4+2592*s^3+136080*s^2+2552384*s+15052095 
\end{colorverbatim}

\section{Conclusions}\label{s:con}

This work has presented $\mathtt{SUGAR}$, a Matlab toolbox for symbolic and numerical computing with geometric algebras. SUGAR complements existing GA software by offering the capacity to perform symbolic computations and providing a mathematical and engineering-oriented design to facilitate its applicability in different disciplines in a user-friendly manner. To this end, apart from introducing the main building blocks and functions of $\mathtt{SUGAR}$, the article provides several examples ranging from standard robotics to iconic rigid body dynamics simulations, and even covers new applications like the use of geometric algebra for the modeling and analysis of power electronics.

As future work, some interesting remarks have arisen from the different tests performed with $\mathtt{SUGAR}$. In particular, $\mathtt{SUGAR}$ has opened the door to the development of closed-form formulas for the inverse of a multivector, the inverse of a matrix of multivectors, and even the exponential of a multivector, which have not been fully addressed in the literature. In addition, other less-used functions like $\log$, $\sin$, $\cos$ can also benefit from the closed-form formulas that can be obtained with $\mathtt{SUGAR}$.



\begin{acks}
The work of I. Zaplana was partially supported by the Spanish National Project PID2020-114819GB-I00 funded by MICIU/AEI/10.13039/501100011033, and by the Generalitat de Catalunya through the Project 2021 SGR 00375. The work of A. Dòria-Cerezo was partially supported by the Spanish National Project PID2021-122821NB-I00 funded by MICIU/AEI/10.13039/501100011033 and ERDF/EU, and by the Generalitat de Catalunya through the Project 2021 SGR 00376. M. Velasco, J. Duarte and P. Martí were partially supported by the Spanish National Project PID2021-122835OB-C21 funded by MICIU/AEI/10.13039/501100011033.
\end{acks}

\bibliographystyle{ACM-Reference-Format}
\bibliography{SuGAR_paper}


\newcommand{\noopsort}[1]{} \newcommand{\printfirst}[2]{#1}
  \newcommand{\singleletter}[1]{#1} \newcommand{\switchargs}[2]{#2#1}
\begin{thebibliography}{35}


\ifx \showCODEN    \undefined \def \showCODEN     #1{\unskip}     \fi
\ifx \showDOI      \undefined \def \showDOI       #1{#1}\fi
\ifx \showISBNx    \undefined \def \showISBNx     #1{\unskip}     \fi
\ifx \showISBNxiii \undefined \def \showISBNxiii  #1{\unskip}     \fi
\ifx \showISSN     \undefined \def \showISSN      #1{\unskip}     \fi
\ifx \showLCCN     \undefined \def \showLCCN      #1{\unskip}     \fi
\ifx \shownote     \undefined \def \shownote      #1{#1}          \fi
\ifx \showarticletitle \undefined \def \showarticletitle #1{#1}   \fi
\ifx \showURL      \undefined \def \showURL       {\relax}        \fi
\providecommand\bibfield[2]{#2}
\providecommand\bibinfo[2]{#2}
\providecommand\natexlab[1]{#1}
\providecommand\showeprint[2][]{arXiv:#2}

\bibitem[Ablamowicz and Fauser(2005)]%
        {Abl05}
\bibfield{author}{\bibinfo{person}{Rafal Ablamowicz} {and}
  \bibinfo{person}{Bertfried Fauser}.} \bibinfo{year}{2005}\natexlab{}.
\newblock \showarticletitle{Mathematics of {CLIFFORD} - A {M}aple package for
  {C}lifford and {G}rassmann algebras}.
\newblock \bibinfo{journal}{\emph{Advances in Applied Clifford Algebras}}
  \bibinfo{volume}{15} (\bibinfo{date}{01} \bibinfo{year}{2005}),
  \bibinfo{pages}{157--181}.
\newblock
\urldef\tempurl%
\url{https://doi.org/10.1007/s00006-005-0009-9}
\showDOI{\tempurl}


\bibitem[Antanovskii(2014)]%
        {Ant14}
\bibfield{author}{\bibinfo{person}{L.K. Antanovskii}.}
  \bibinfo{year}{2014}\natexlab{}.
\newblock \bibinfo{booktitle}{\emph{Implementation of Geometric Algebra in
  {MATLAB} with Applications}}.
\newblock \bibinfo{type}{{T}echnical {R}eport}.
  \bibinfo{institution}{DSTO–TR–3021 Weapons and Combat Systems Division,
  Defence Science and Technology Organisation, Dept. of Defense, Australian
  Government},
  \bibinfo{address}{\url{https://apps.dtic.mil/sti/pdfs/ADA615302.pdf}}.
\newblock


\bibitem[Awad~Eid(2016)]%
        {Eid16}
\bibfield{author}{\bibinfo{person}{Ahmad~Hosney Awad~Eid}.}
  \bibinfo{year}{2016}\natexlab{}.
\newblock \bibinfo{title}{Optimized Automatic Code Generation for Geometric
  Algebra Based Algorithms with Ray Tracing Application}.
\newblock
\newblock
\showeprint[arxiv]{1607.04767}


\bibitem[Bayro-Corrochano(2020)]%
        {Bay20}
\bibfield{author}{\bibinfo{person}{Eduardo Bayro-Corrochano}.}
  \bibinfo{year}{2020}\natexlab{}.
\newblock \bibinfo{booktitle}{\emph{Geometric Algebra Applications Vol. II:
  Robot Modelling and Control}}.
\newblock \bibinfo{publisher}{Springer}.
\newblock
\showISBNx{978-3-030-34976-9}
\urldef\tempurl%
\url{https://doi.org/10.1007/978-3-030-34978-3}
\showDOI{\tempurl}


\bibitem[Bayro-Corrochano(2021)]%
        {Bay21}
\bibfield{author}{\bibinfo{person}{Eduardo Bayro-Corrochano}.}
  \bibinfo{year}{2021}\natexlab{}.
\newblock \showarticletitle{A Survey on Quaternion Algebra and Geometric
  Algebra Applications in Engineering and Computer Science 1995–2020}.
\newblock \bibinfo{journal}{\emph{IEEE Access}}  \bibinfo{volume}{9}
  (\bibinfo{year}{2021}), \bibinfo{pages}{104326--104355}.
\newblock
\urldef\tempurl%
\url{https://doi.org/10.1109/ACCESS.2021.3097756}
\showDOI{\tempurl}


\bibitem[Bayro-Corrochano et~al\mbox{.}(2022)]%
        {Bay22}
\bibfield{author}{\bibinfo{person}{Eduardo Bayro-Corrochano},
  \bibinfo{person}{Jesus Medrano-Hermosillo}, \bibinfo{person}{Guillermo
  Osuna-González}, {and} \bibinfo{person}{Ulises Uriostegui-Legorreta}.}
  \bibinfo{year}{2022}\natexlab{}.
\newblock \showarticletitle{Newton–Euler modeling and Hamiltonians for robot
  control in the geometric algebra}.
\newblock \bibinfo{journal}{\emph{Robotica}} \bibinfo{volume}{40},
  \bibinfo{number}{11} (\bibinfo{year}{2022}), \bibinfo{pages}{4031–4055}.
\newblock
\urldef\tempurl%
\url{https://doi.org/10.1017/S0263574722000741}
\showDOI{\tempurl}


\bibitem[Bromborsky and team(2014)]%
        {galgebra}
\bibfield{author}{\bibinfo{person}{Alan Bromborsky} {and}
  \bibinfo{person}{GAlgebra team}.} \bibinfo{year}{2014}\natexlab{}.
\newblock \bibinfo{booktitle}{\emph{GAlgebra}}.
\newblock Army Research Lab.
\newblock
\urldef\tempurl%
\url{https://github.com/pygae/galgebra}
\showURL{%
\tempurl}


\bibitem[Calvet(2017)]%
        {Cal17}
\bibfield{author}{\bibinfo{person}{Ramon Calvet}.}
  \bibinfo{year}{2017}\natexlab{}.
\newblock \showarticletitle{On Matrix Representations of Geometric (Clifford)
  Algebras}.
\newblock \bibinfo{journal}{\emph{Journal of Geometry and Symmetry in Physics}}
   \bibinfo{volume}{43} (\bibinfo{date}{01} \bibinfo{year}{2017}),
  \bibinfo{pages}{1--36}.
\newblock
\urldef\tempurl%
\url{https://doi.org/10.7546/jgsp-43-2017-1-36}
\showDOI{\tempurl}


\bibitem[Chappell et~al\mbox{.}(2014)]%
        {Cha14}
\bibfield{author}{\bibinfo{person}{James~M. Chappell},
  \bibinfo{person}{Samuel~P. Drake}, \bibinfo{person}{Cameron~L. Seidel},
  \bibinfo{person}{Lachlan~J. Gunn}, \bibinfo{person}{Azhar Iqbal},
  \bibinfo{person}{Andrew Allison}, {and} \bibinfo{person}{Derek Abbott}.}
  \bibinfo{year}{2014}\natexlab{}.
\newblock \showarticletitle{Geometric Algebra for Electrical and Electronic
  Engineers}.
\newblock \bibinfo{journal}{\emph{Proc. IEEE}} \bibinfo{volume}{102},
  \bibinfo{number}{9} (\bibinfo{year}{2014}), \bibinfo{pages}{1340--1363}.
\newblock
\urldef\tempurl%
\url{https://doi.org/10.1109/JPROC.2014.2339299}
\showDOI{\tempurl}


\bibitem[Colapinto(2011)]%
        {Ver11}
\bibfield{author}{\bibinfo{person}{Pablo Colapinto}.}
  \bibinfo{year}{2011}\natexlab{}.
\newblock \emph{\bibinfo{title}{Versor: Spatial Computing with Conformal
  Geometric Algebra}}.
\newblock \bibinfo{thesistype}{Master's\ thesis}. \bibinfo{school}{University
  of California Santa Barbara},
  \bibinfo{address}{\url{http://versor.mat.ucsb.edu}}.
\newblock
\urldef\tempurl%
\url{http://versor.mat.ucsb.edu}
\showURL{%
\tempurl}


\bibitem[Dargys and Acus(2021)]%
        {exponentials}
\bibfield{author}{\bibinfo{person}{Adolfas Dargys} {and}
  \bibinfo{person}{Arturas Acus}.} \bibinfo{year}{2021}\natexlab{}.
\newblock \bibinfo{title}{Exponentials of general multivector (MV) in 3D
  Clifford algebras}.
\newblock
\newblock
\showeprint[arxiv]{2104.01905}


\bibitem[De~Keninck(2017)]%
        {GANJA}
\bibfield{author}{\bibinfo{person}{Steven De~Keninck}.}
  \bibinfo{year}{2017}\natexlab{}.
\newblock \bibinfo{booktitle}{\emph{ganja.js}}.
\newblock University of Amsterdam.
\newblock
\urldef\tempurl%
\url{https://github.com/enkimute/ganja.js}
\showURL{%
\tempurl}


\bibitem[Dimo(1975)]%
        {Dimo1975}
\bibfield{author}{\bibinfo{person}{Paul Dimo}.}
  \bibinfo{year}{1975}\natexlab{}.
\newblock \bibinfo{booktitle}{\emph{Nodal analysis of power systems}}.
\newblock \bibinfo{publisher}{Abacus Press}.
\newblock


\bibitem[Doran and Lasenby(2003)]%
        {Doran03}
\bibfield{author}{\bibinfo{person}{C. Doran} {and} \bibinfo{person}{A.
  Lasenby}.} \bibinfo{year}{2003}\natexlab{}.
\newblock \bibinfo{booktitle}{\emph{Geometric Algebra for Physicists}}.
\newblock \bibinfo{publisher}{Cambridge University Press}.
\newblock


\bibitem[Dorst and Keninck(2022)]%
        {Dor22}
\bibfield{author}{\bibinfo{person}{Leo Dorst} {and} \bibinfo{person}{Steven~De
  Keninck}.} \bibinfo{year}{2022}\natexlab{}.
\newblock \bibinfo{title}{Guided tour to the plane-based geometric algebra
  PGA}.
\newblock \bibinfo{howpublished}{\url{https://bivector.net/PGA4CS.html}}.
\newblock


\bibitem[Dòria-Cerezo et~al\mbox{.}(2024)]%
        {DVZM2024}
\bibfield{author}{\bibinfo{person}{Arnau Dòria-Cerezo}, \bibinfo{person}{Manel
  Velasco}, \bibinfo{person}{Isiah Zaplana}, {and} \bibinfo{person}{Pau
  Martí}.} \bibinfo{year}{2024}\natexlab{}.
\newblock \showarticletitle{{The Ohm’s law for (non-symmetric) three-phase
  three-wire electrical circuits: from a complex-valued to a geometric algebra
  approach}}.
\newblock In \bibinfo{booktitle}{\emph{Proc. International Conference on
  Modeling and Simulation of Electric Machines, Converters and Systems}}.
\newblock


\bibitem[Hestenes(2001)]%
        {Hes01}
\bibfield{author}{\bibinfo{person}{David Hestenes}.}
  \bibinfo{year}{2001}\natexlab{}.
\newblock \showarticletitle{Old Wine in New Bottles: A New Algebraic Framework
  for Computational Geometry}.
\newblock In \bibinfo{booktitle}{\emph{Geometric Algebra with Applications in
  Science and Engineering}}, \bibfield{editor}{\bibinfo{person}{Eduardo~Bayro
  Corrochano} {and} \bibinfo{person}{Garret Sobczyk}} (Eds.).
  \bibinfo{publisher}{Birkh{\"a}user Boston}, \bibinfo{pages}{3--17}.
\newblock


\bibitem[Hestenes and Sobczyk(1984)]%
        {Hes84}
\bibfield{author}{\bibinfo{person}{David Hestenes} {and}
  \bibinfo{person}{Garret Sobczyk}.} \bibinfo{year}{1984}\natexlab{}.
\newblock \bibinfo{booktitle}{\emph{Clifford algebra to geometric calculus : a
  unified language for mathematics and physics}}.
\newblock \bibinfo{publisher}{D. Reidel ; Distributed in the U.S.A. and Canada
  by Kluwer Academic Publishers}, \bibinfo{address}{Dordrecht; Boston; Hingham,
  MA, U.S.A.}
\newblock
\showISBNx{9027716730 9789027716736 9027725616 9789027725615}


\bibitem[Hitzer et~al\mbox{.}(2024)]%
        {Hit24}
\bibfield{author}{\bibinfo{person}{Eckhard Hitzer}, \bibinfo{person}{Carlile
  Lavor}, {and} \bibinfo{person}{Dietmar Hildenbrand}.}
  \bibinfo{year}{2024}\natexlab{}.
\newblock \showarticletitle{Current survey of Clifford geometric algebra
  applications}.
\newblock \bibinfo{journal}{\emph{Mathematical Methods in the Applied
  Sciences}} \bibinfo{volume}{47}, \bibinfo{number}{3} (\bibinfo{year}{2024}),
  \bibinfo{pages}{1331--1361}.
\newblock
\urldef\tempurl%
\url{https://doi.org/10.1002/mma.8316}
\showDOI{\tempurl}


\bibitem[Hitzer et~al\mbox{.}(2013)]%
        {Hit13}
\bibfield{author}{\bibinfo{person}{Eckhard Hitzer}, \bibinfo{person}{Tohru
  Nitta}, {and} \bibinfo{person}{Yasuaki Kuroe}.}
  \bibinfo{year}{2013}\natexlab{}.
\newblock \showarticletitle{Applications of Clifford's Geometric Algebra}.
\newblock \bibinfo{journal}{\emph{Advances in Applied Clifford Algebras}}
  \bibinfo{volume}{23}, \bibinfo{number}{2} (\bibinfo{year}{2013}),
  \bibinfo{pages}{377--404}.
\newblock
\urldef\tempurl%
\url{https://doi.org/10.1007/s00006-013-0378-4}
\showDOI{\tempurl}


\bibitem[Hitzer and Sangwine(2017)]%
        {Hitzer_inv}
\bibfield{author}{\bibinfo{person}{Eckhard Hitzer} {and}
  \bibinfo{person}{Stephen Sangwine}.} \bibinfo{year}{2017}\natexlab{}.
\newblock \showarticletitle{Multivector and multivector matrix inverses in real
  Clifford algebras}.
\newblock \bibinfo{journal}{\emph{Appl. Math. Comput.}}  \bibinfo{volume}{311}
  (\bibinfo{year}{2017}), \bibinfo{pages}{375--389}.
\newblock
\showISSN{0096-3003}
\urldef\tempurl%
\url{https://doi.org/10.1016/j.amc.2017.05.027}
\showDOI{\tempurl}


\bibitem[Lavor et~al\mbox{.}(2018)]%
        {Lav18}
\bibfield{author}{\bibinfo{person}{C. Lavor}, \bibinfo{person}{S.
  Xambó-Descamps}, {and} \bibinfo{person}{I. Zaplana}.}
  \bibinfo{year}{2018}\natexlab{}.
\newblock \bibinfo{booktitle}{\emph{A Geometric Algebra Invitation to
  Space-Time Physics, Robotics and Molecular Geometry}}.
\newblock \bibinfo{publisher}{SpringerBriefs in Mathematics, Springer Cham}.
\newblock
\showISBNx{978-3-319-90664-5}


\bibitem[Lounesto et~al\mbox{.}(1987)]%
        {Lou87}
\bibfield{author}{\bibinfo{person}{Pertti Lounesto}, \bibinfo{person}{R.
  Mikkola}, {and} \bibinfo{person}{V. Vierros}.}
  \bibinfo{year}{1987}\natexlab{}.
\newblock \bibinfo{booktitle}{\emph{CLICAL user manual: complex number, vector
  space and Clifford algebra calculator for MS-DOS personal computers}}.
\newblock \bibinfo{type}{{T}echnical {R}eport}. \bibinfo{institution}{Helsinki
  University of Technology, Finland},
  \bibinfo{address}{\url{https://users.aalto.fi/~ppuska/mirror/Lounesto/CLICAL.htm}}.
\newblock


\bibitem[L{\"o}w and Calinon(2023)]%
        {Tob23}
\bibfield{author}{\bibinfo{person}{Tobias L{\"o}w} {and}
  \bibinfo{person}{Sylvain Calinon}.} \bibinfo{year}{2023}\natexlab{}.
\newblock \showarticletitle{Geometric Algebra for Optimal Control With
  Applications in Manipulation Tasks}.
\newblock \bibinfo{journal}{\emph{IEEE Transactions on Robotics}}
  \bibinfo{volume}{39}, \bibinfo{number}{5} (\bibinfo{year}{2023}),
  \bibinfo{pages}{3586--3600}.
\newblock
\urldef\tempurl%
\url{https://doi.org/10.1109/TRO.2023.3277282}
\showDOI{\tempurl}


\bibitem[Mann et~al\mbox{.}(2001)]%
        {Mann99}
\bibfield{author}{\bibinfo{person}{Stephen Mann}, \bibinfo{person}{Leo Dorst},
  {and} \bibinfo{person}{Tim Bouma}.} \bibinfo{year}{2001}\natexlab{}.
\newblock \showarticletitle{{The Making of GABLE: A Geometric Algebra Learning
  Environment in Matlab}}.
\newblock In \bibinfo{booktitle}{\emph{Geometric Algebra with Applications in
  Science and Engineering}}, \bibfield{editor}{\bibinfo{person}{Eduardo~Bayro
  Corrochano} {and} \bibinfo{person}{Garret Sobczyk}} (Eds.).
  \bibinfo{publisher}{Birkh{\"a}user Boston}, \bibinfo{address}{Boston, MA},
  \bibinfo{pages}{491--511}.
\newblock
\urldef\tempurl%
\url{https://doi.org/10.1007/978-1-4612-0159-5_24}
\showDOI{\tempurl}


\bibitem[Ortiz-Duran and Aragon(2017)]%
        {Ara22}
\bibfield{author}{\bibinfo{person}{E.~Alejandra Ortiz-Duran} {and}
  \bibinfo{person}{Jose~L. Aragon}.} \bibinfo{year}{2017}\natexlab{}.
\newblock \bibinfo{title}{CGAlgebra: a Mathematica package for conformal
  geometric algebra. v.2.0}.
\newblock
\newblock
\showeprint[arxiv]{1711.02513}


\bibitem[Prodanov and Toth(2017)]%
        {Pro16}
\bibfield{author}{\bibinfo{person}{Dimiter Prodanov} {and} \bibinfo{person}{V.
  Toth}.} \bibinfo{year}{2017}\natexlab{}.
\newblock \showarticletitle{Sparse Representations of Clifford and Tensor
  Algebras in {M}axima}.
\newblock \bibinfo{journal}{\emph{Advances in Applied Clifford Algebras}}
  \bibinfo{volume}{27} (\bibinfo{year}{2017}), \bibinfo{pages}{661--683}.
\newblock
\urldef\tempurl%
\url{https://doi.org/10.1007/s00006-016-0682-x}
\showDOI{\tempurl}


\bibitem[Roelfs and Keninck(2023)]%
        {Roe21}
\bibfield{author}{\bibinfo{person}{Martin Roelfs} {and}
  \bibinfo{person}{Steven~De Keninck}.} \bibinfo{year}{2023}\natexlab{}.
\newblock \showarticletitle{Graded Symmetry Groups: Plane and Simple}.
\newblock \bibinfo{journal}{\emph{Advances in Applied Clifford Algebras}}
  \bibinfo{volume}{30}, \bibinfo{number}{33} (\bibinfo{year}{2023}),
  \bibinfo{pages}{1--41}.
\newblock
\urldef\tempurl%
\url{https://doi.org/10.1007/s00006-023-01269-9}
\showDOI{\tempurl}


\bibitem[Sangwine and Le~Bihan(2005)]%
        {qtfm}
\bibfield{author}{\bibinfo{person}{Steve Sangwine} {and}
  \bibinfo{person}{Nicolas Le~Bihan}.} \bibinfo{year}{2005}\natexlab{}.
\newblock \bibinfo{booktitle}{\emph{Quaternion and octonion toolbox for
  Matlab}}.
\newblock Grenoble-INP and Université de Grenoble-Alpes.
\newblock
\urldef\tempurl%
\url{https://qtfm.sourceforge.io/}
\showURL{%
\tempurl}


\bibitem[Sangwine and Hitzer(2016)]%
        {San16}
\bibfield{author}{\bibinfo{person}{Stephen~J. Sangwine} {and}
  \bibinfo{person}{Eckhard Hitzer}.} \bibinfo{year}{2016}\natexlab{}.
\newblock \showarticletitle{Clifford Multivector Toolbox (for {MATLAB})}.
\newblock \bibinfo{journal}{\emph{Advances in Applied Clifford Algebras}}
  \bibinfo{volume}{27}, \bibinfo{number}{1} (\bibinfo{date}{apr}
  \bibinfo{year}{2016}), \bibinfo{pages}{539--558}.
\newblock
\urldef\tempurl%
\url{https://doi.org/10.1007/s00006-016-0666-x}
\showDOI{\tempurl}


\bibitem[Siciliano et~al\mbox{.}(2008)]%
        {Siciliano08a}
\bibfield{author}{\bibinfo{person}{B. Siciliano}, \bibinfo{person}{L.
  Sciavicco}, \bibinfo{person}{L. Villani}, {and} \bibinfo{person}{G. Oriolo}.}
  \bibinfo{year}{2008}\natexlab{}.
\newblock \bibinfo{booktitle}{\emph{Robotics: Modelling, Planning and
  Control}}.
\newblock \bibinfo{publisher}{Springer Publishing Company}.
\newblock


\bibitem[Velasco(2023)]%
        {SUGAR}
\bibfield{author}{\bibinfo{person}{Manel Velasco}.}
  \bibinfo{year}{2023}\natexlab{}.
\newblock \bibinfo{booktitle}{\emph{Symbolic and User-friendly Geometric
  Algebra Routines (SUGAR)}}.
\newblock Universitat Politècnica de Catalunya (UPC).
\newblock
\urldef\tempurl%
\url{https://github.com/distributed-control-systems/SUGAR}
\showURL{%
\tempurl}


\bibitem[Velasco et~al\mbox{.}(2025)]%
        {Vel23}
\bibfield{author}{\bibinfo{person}{Manel Velasco}, \bibinfo{person}{Isiah
  Zaplana}, \bibinfo{person}{Arnau D{\`o}ria-Cerezo}, \bibinfo{person}{Josu{\'
  e} Duarte}, {and} \bibinfo{person}{Pau Mart{\' i}}.}
  \bibinfo{year}{2025}\natexlab{}.
\newblock \showarticletitle{Introducing Modelling, Analysis and Control of
  Three-Phase Electrical Systems Using Geometric Algebra}.
\newblock \bibinfo{journal}{\emph{IEEE Transactions on Industrial Electronics}}
   \bibinfo{volume}{Accepted for publication} (\bibinfo{year}{2025}),
  \bibinfo{pages}{1--10}.
\newblock
\urldef\tempurl%
\url{https://doi.org/10.1109/TIE.2025.3553169}
\showDOI{\tempurl}


\bibitem[Zaplana et~al\mbox{.}(2022a)]%
        {Zap21}
\bibfield{author}{\bibinfo{person}{Isiah Zaplana}, \bibinfo{person}{Hugo
  Hadfield}, {and} \bibinfo{person}{Joan Lasenby}.}
  \bibinfo{year}{2022}\natexlab{a}.
\newblock \showarticletitle{Closed-form solutions for the inverse kinematics of
  serial robots using conformal geometric algebra}.
\newblock \bibinfo{journal}{\emph{Mechanism and Machine Theory}}
  \bibinfo{volume}{173} (\bibinfo{year}{2022}), \bibinfo{pages}{104835}.
\newblock
\showISSN{0094-114X}
\urldef\tempurl%
\url{https://doi.org/10.1016/j.mechmachtheory.2022.104835}
\showDOI{\tempurl}


\bibitem[Zaplana et~al\mbox{.}(2022b)]%
        {Isi22}
\bibfield{author}{\bibinfo{person}{Isiah Zaplana}, \bibinfo{person}{Hugo
  Hadfield}, {and} \bibinfo{person}{Joan Lasenby}.}
  \bibinfo{year}{2022}\natexlab{b}.
\newblock \showarticletitle{Singularities of Serial Robots: Identification and
  Distance Computation Using Geometric Algebra}.
\newblock \bibinfo{journal}{\emph{Mathematics}} \bibinfo{volume}{10},
  \bibinfo{number}{12} (\bibinfo{year}{2022}), \bibinfo{pages}{2068}.
\newblock
\showISSN{2227-7390}
\urldef\tempurl%
\url{https://doi.org/10.3390/math10122068}
\showDOI{\tempurl}


\end{thebibliography}



%

\end{document}